\documentclass[fleqn,usenatbib]{mnras}

\usepackage[T1]{fontenc}
\usepackage{ae,aecompl}

\usepackage{graphicx}	
\usepackage{amsmath}	
\usepackage{amssymb}	
\usepackage[normalem]{ulem}
\usepackage{graphicx,color}	

\newcommand{\kms}{\textrm{km s}^{-1}}

\newcommand{\msol}{\text{M}$_{\odot}$}
\newcommand{\pc}{\text{pc}}
\newcommand{\Myr}{\text{Myr}}
\newcommand{\yr}{\text{yr}}
\newcommand{\kpc}{\mbox{$\>{\rm kpc}$}}

\newcommand{\galactics}{\textsc{GalactICS}}

\title[Milky Way BP Bulge]{Simulating the Milky Way bar and bulge with an initially S\'ersic disc}

\author[N. Deg et al.]{
N. Deg$^{1}$,
Victor P. Debattista$^{2}$\thanks{E-mail: vpdebattista@gmail.com},
Lawrence Widrow$^{1}$,
Stuart Robert Anderson$^{2}$, \newauthor
Oscar A. Gonzalez$^3$ and
Thomas R. Quinn$^4$
\\
$^{1}$ Department of Physics, Engineering Physics, and Astronomy, Queen's University, Kingston, ON, K7L 3N6, Canada\\
$^2$ Jeremiah Horrocks Institute, University of Central Lancashire, Preston PR1 2HE, UK\\
$^3$ UK Astronomy Technology Centre, Royal Observatory, Blackford Hill, Edinburgh EH9 3HJ, UK \\
$^4$ Astronomy Department, University of Washington, Box 351580, Seattle, WA 98195, USA
}

\date{Accepted XXX. Received YYY; in original form ZZZ}

\pubyear{2024}

\begin{document}
\label{firstpage}
\pagerange{\pageref{firstpage}--\pageref{lastpage}}
\maketitle

\begin{abstract}
We model the formation of a bar plus box/peanut bulge (BP bulge) component in a Milky Way-like disc galaxy using simulations of isolated multi-component systems that evolve from equilibrium initial conditions. The simulations are designed to test the hypothesis that the bar forms early on and thickens to create the bulge. To this end, our initial conditions include a stellar disc with a S\'{e}rsic surface density profile and do not include any classical bulge component. We also include a gas disc, which is important in regulating the growth of the bar. Our best-fit model has an initial stellar disc with a Sersic index of $n = 1.75$ and a gas disc with mass equal to 7\% of the mass of the stellar disc. The model reproduces the bar size, pattern speed, and box/peanut shape of the Milky Way's bulge+bar.

\end{abstract}

\begin{keywords}
Galaxy: kinematics and dynamics -- Galaxy: structure -- methods: numerical
\end{keywords}



\section{Introduction}

The field of Galactic astronomy grew, in large part, out of attempts to understand the formation, structure, and evolution of the Milky Way (MW). Models of the Galaxy can vary in scope and level of detail, depending on the questions that one intends to explore. At one end, kinematic models for the Galaxy's stellar components provide the phase space distribution function (DF) of the stars without regard for the gravitational potential (for some examples see \citealt{Juric2008,Bond2010,Binney2010}). By contrast, equilibrium dynamical models include the DFs for all massive components (stars, gas, dark matter) as well as a self-consistent model for the gravitational potential under the assumption that the system is stationary \citep{Kuijken1995,Robin2003,Widrow2005,Binney2012,McMillan2017,Vasiliev2019,Binney2023}. In general, equilibrium models are symmetric about the spin axis of the Galaxy and its mid-plane and therefore cannot account for the Galaxy's bar, spiral arms, or warp. However, one can explore the formation of non-equilibrium structures such as these by evolving equilibrium models using N-body methods.  This strategy, which dates back to the pioneering work of \citet{Miller1970}, \citet{Ostriker1973}, and others, exploits the fact that equilibrium models are generally susceptible to global and local instabilities, which can drive the formation of a central bar and spiral arms (for example see Ch. 6 of \citealt{Binney2008} and references therein).  The question then is whether the instabilities in the initial system lead it to evolve to a state that is consistent with present-day observations.

As with most disc galaxies, the surface brightness profile of the MW rises above an exponential near the centre. This excess light is often attributed to a central bulge. Simulations of MW-like galaxies typically model this through the inclusion of a slowly rotating classical bulge \citep[e.g.][]{fujii+19, D'Onghia2020, TepperGarcia2021}, that is, a dynamically distinct, centrally concentrated component, into their initial conditions (ICs).  However, there is compelling evidence that the MW has a box/peanut-shaped (BP) bulge, that is, a rotationally supported stellar component that formed through secular processes involving the thickening of the bar \citep{Kormendy2004,Shen2010,Debattista2017,KormendyBender2019}.  The appropriate ICs to test this hypothesis therefore must be a bulgeless disc galaxy.

Simulations of a MW-like galaxy must also reproduce the length, strength, and pattern speed of the Galactic bar.  Bar formation proceeds within a resonant cavity of spiral density waves which reflect between the centre and the corotation resonance \citep{Toomre1972}. The resulting bars extend up to the largest radius corresponding to the corotation of the slowest spiral that avoids an inner Lindblad resonance (ILR). A higher central mass concentration, such as that resulting from a bulge, raises the ILR curve, which means that the spiral that can avoid an ILR must be faster, and therefore the resonant cavity smaller, which results in a smaller bar forming \citep{Toomre1972, Sellwood1985}. Once formed, bars evolve by shedding angular momentum to the dark matter halo. As they do so, their pattern speed decreases and they grow in length \citep[e.g.][]{debattista_sellwood00, Athanassoula2003, oneill_dubinski+03, holley-bockelmann_weinberg05, martinez-valpuesta+06, weinberg_katz07, sellwood2016bar, polyachenko2016, kataria2022effects, Joshi2024}. This process is very efficient when the halo is centrally concentrated, which is the case for NFW halos. The rapid growth of the bar in models with cuspy halos can make it difficult to simultaneously satisfy observational constraints on both the length and pattern speed. For example, the bar in the model of \citet{TepperGarcia2021} was able to match the pattern speed of the MW only for a brief time interval, $\sim 2-3$~Gyr (depending on the specific contraint selected). Not only does this rapid growth produce bars which are too large compared with observations \citep{erwin2005}, but the high slowdown rate reduces the efficiency at which resonances can trap stars \citep[e.g.][]{Weinberg85, Chiba2021}. 

It is possible that the failure of simulations to reproduce the observed pattern speed and length of the bar is due to an incorrect model of the dark halo since both the growth and spindown of the bar are driven, to a large extent, by a transfer of angular momentum from the disc to the halo. \citet{Athanassoula2003} showed that bar formation was less vigorous in a static halo, while work by a variety of different groups demonstrated that the growth and evolution of bar depended on whether the halo was rotating with or counter to the disc. \citep{debattista_sellwood00, fujii+19,collier2019, kataria2022effects,chiba_kataria24}.

Though most studies of bar formation in simulations of isolated disc galaxies have included only a stellar (i.e., collisionless) disc, the presence of a gas disc may be important for regulating bar growth.  \citet{Lokas2020} note that gas discs may weaken the bar instability. \citet{Beane2023} argued that gas-rich galaxies have bars that do not slow down, which they interpreted as resulting from a steady supply of angular momentum from the gas to the bar. They relate this mechanism to the metastability discussed by \citet{sellwood_debattista06}, in which any process which causes the bar pattern speed to increase briefly (such as a sudden increase in the central density) gives rise to resonances facing a rising phase space density, which inhibits slowdown for a long time. \citet{sellwood_debattista06} emphasize that this metastable state is quite sensitive to small perturbations. However, in an isolated galaxy with no interactions, a bar can persist in the metastable state for several gigayears.  

In this work, we consider the evolution of bulgeless disc-halo systems using initial discs that are more centrally concentrated than pure exponential discs. Other examples of disc models with dense cores can be found in \citet{evans1998} and  \citet{Jalali2005}.  Here we consider discs with surface density profiles given by a S\'{e}rsic profile, $\Sigma(R)\propto e^{-(R/R_d)^{1/n}}$, which is a generalization of the exponential disc. For $n>1$, S\'{e}rsic discs have an excess mass at small radii as compared to exponential discs. 

The models are built using a modified version of the Galaxy Initial ConditionS code \citep[\galactics;][]{Kuijken1995,Widrow2005,Widrow2008,Deg2019}.  \galactics\ is designed to generate multi-component equilibrium ICs for $N$-body simulations of galaxies.  Previous versions of the code build stellar discs with exponential surface density profiles, but our publicly available version of the code\footnote{\href{https://github.com/NateDeg/GalactICS_SersicDisk.git}{https://github.com/NateDeg/GalactICS\_SersicDisk.git}} builds S\'{e}rsic stellar discs.

This paper is organised as follows. We motivate the use of S\'ersic discs in Section~\ref{Sec:Sersic}.
In Section~\ref{Sec:ICs} we describe our ICs including the implementation of the S\'{e}rsic disc, as well as details of the simulations. Section~\ref{Sec:GasAndBars} presents the evolution of both the bar and the BP-bulge in these initially bulgeless systems. We summarize our results in Section~\ref{Sec:Conclusions}.  Appendix~\ref{sec:Galactics} provides a more detailed description of the \galactics\ modifications necessary to build S\'{e}rsic discs, while Appendix~\ref{MWComp} presents additional comparisons of our MW-like simulation to observations of the Galaxy.


\section{Motivation for S\'ersic Discs}
\label{Sec:Sersic}

Though the exponential disc was originally proposed as an empirical fit to observational data \citep{Freeman1970} there have been various attempts to motivate it from first principles. For example \citet{Fall1980} and \citet{Mo1998} showed that an exponential disc can arise from a primordial rotating gas sphere under the assumption that the specific angular momentum of the gas is conserved as it collapses to a rotationally supported disc. These and other arguments provide a strong plausibility argument for approximately exponential discs but do not preclude departures from a pure exponential profile. For example, \citet{Herpich2017} provide a theoretical argument for a surface density profile that deviates from a pure exponential in a manner that depends on the shape of the circular speed curve. Their argument is that, from a maximal entropy principle, the angular momentum should follow an exponential profile.  In such a maximal entropy disc (MED), radial migration scrambles the angular momentum of individual stars while conserving the total mass and angular momentum of the system, leading the specific angular momentum distribution, $N(j)$, to be
\begin{equation}
  dN \propto e^{-j/\langle j\rangle } dj
\end{equation}
where $\langle j\rangle$ is a constant. This translates to a surface brightness profile
\begin{equation}
  \Sigma(R) \propto \frac{v_c(R)}{\langle j\rangle R} \left (
    1 + \frac{d\log v_c(R)}{d\log{R}}\right ) e^{-Rv_c(R)/\langle
    j\rangle}
  \end{equation}
where $v_c$ is the circular speed. For a flat rotation curve, $\Sigma\propto e^{-R/R_d}/R$ where $R_d = \langle j\rangle/v_c$. Thus, the model predicts an exponential profile for $R\gg R_e$ but one that rises above a pure exponential at smaller radii. On the other hand, for solid body rotation ($v = \Omega R$), one has $\Sigma\propto e^{-R^2/R_e^2}$ where $R_e = \sqrt{\langle j\rangle/\Omega}$.

The MED has a striking similarity to a S\'{e}rsic disc, which has a surface density profile of
\begin{equation}
  \Sigma(R) = \Sigma_0 e^{-(R/R_d)^{1/n}},
\end{equation}
and a total mass of
\begin{equation}
  M = 2^{2n}\sqrt{\pi}\Sigma_0 R_d^2 \Gamma(n + 1)\Gamma(n + \frac12)~,
\end{equation}
where $\Sigma_{0}$ is the central surface density, $R_{d}$ is the radial scale length, $n$ is the S\'{e}rsic index, and $\Gamma$ is the Gamma function.  When $n=1$, the S\'{e}rsic disc reduces to an exponential disc. For $n>1$, the S\'{e}rsic surface density profile rises above the exponential profile in its inner regions, similar to MEDs with flat rotation curves. On the other hand the S\'{e}rsic disc with $n=1/2$ corresponds to a MED with solid body rotation. Note that the interpretation of the radial scale length depends on $n$. For example, the mass weighted average of $1/R$ (appropriate as an estimator of the potential) is
\begin{equation}
  \left\langle{R^{-1}}\right\rangle  = 
  2^{2n-1}\pi^{-1/2}\Gamma(n+1/2) R_d^{-1}
\end{equation}
which equals $1,\, 0.443,\,0.167~ R_d^{-1}$ for $n=1,\,1.5,\,2$, respectively.

S\'ersic disc profiles were suggested by \citet{boeker+03} for late-type galaxies and \citet{Debattista2006} presented examples of $N$-body simulations of barred disc galaxies with initial S\'ersic discs (up to $n=2.5$) embedded in unresponsive halo potentials.

\begin{figure*}
\centering
    \includegraphics[width=0.8\textwidth]{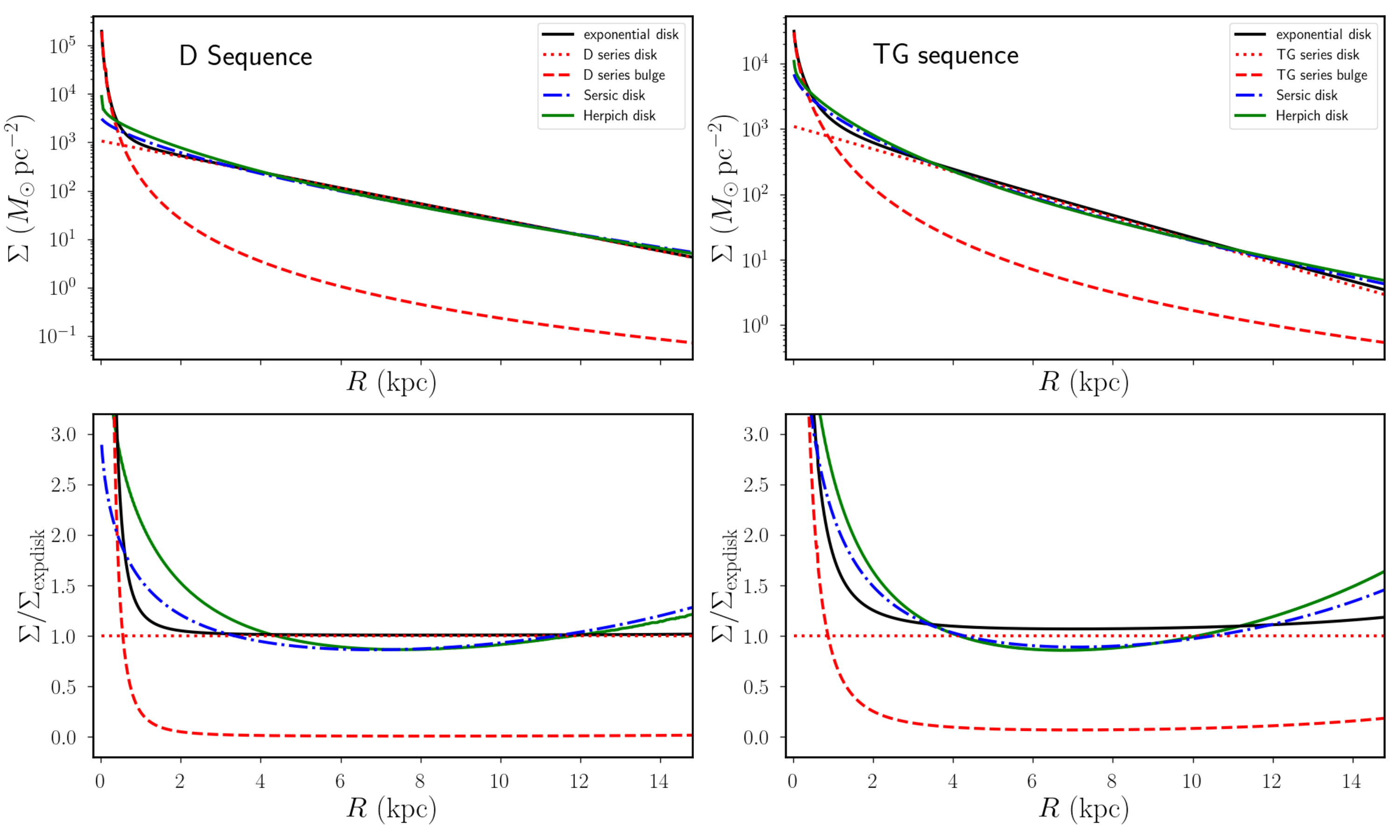}
    \caption{Surface density profiles for bulge$+$disc, S\'{e}rsic, and MED models. The surface density $\Sigma$ is shown in the top panels on a semi-log plot. The exponential disc (red dotted curves) and bulge (red dashed curves) are shown for the \citet{D'Onghia2020} (left) and \citet{TepperGarcia2021} (right) models. The total surface brightness profiles for these bulge$+$disc models is shown as solid black curves. The S\'{e}rsic models used in this paper are shown as dot-dashed (blue) curves. In the bottom panel, we show the same surface brightness profiles normalised by the exponential disc on a linear plot. A MED that best approximates the bulge$+$disc model is shown as a solid green curve. The models with gas include an additional exponential disc with properties given in Table~\ref{Tab:ModelParams}.}
  \label{Fig:Model_disc_SDComps}
\end{figure*}

Both MEDs and S\'{e}rsic discs can have an enhanced central surface density that mimics the surface density profile of a bulge plus exponential disc model. In Figure~\ref{Fig:Model_disc_SDComps} we compare the surface density profiles of the bulge+disc models of \citet{D'Onghia2020} and \citet{TepperGarcia2021} with comparable surface density profiles that assume either an exponential disc, a S\'{e}rsic disc or an MED (see Sec.~\ref{ssec:ModelDetails} for more details on the  \citet{D'Onghia2020} and \citet{TepperGarcia2021} models). Overall both the MED and S\'{e}rsic disc profiles follow the bulge+disc models better than the pure exponential disc.  They show an excess density at large radii, and they do not rise as quickly as the bulge+disc models in the innermost region.  Nonetheless, both the MED and S\'{e}rsic disc profiles provide reasonable fits for the \citet{D'Onghia2020} and \citet{TepperGarcia2021} surface densities.

\section{Simulation Suite}
\label{Sec:ICs}

We explore the evolution of MW models using bulgeless ICs through a sequence of $N$-body simulations based around two recent, notable high-resolution MW simulations: the \citet{D'Onghia2020} MW model and the \citet{TepperGarcia2021} MW model.  The ICs for our suite of simulations are built using a modification of the \galactics\ code \citep{Kuijken1995,Widrow2005,Widrow2008,Deg2019}.  The \citet{Deg2019} version of the code can generate models with up to five components: an exponential gas disc, two exponential stellar discs, a centrally concentrated bulge, and a double-power-law dark matter halo. To build our suite of ICs we further modify \galactics\ to generate S\'{e}rsic stellar discs.

In the original version of \galactics\ \citep{Kuijken1995}, the DFs for the three components are elementary functions of the energy, $E$, angular momentum momentum about the spin axis of the disc, $L_z$, and the energy of vertical oscillations, $E_z$. The latter is conserved to a good approximation for nearly circular orbits, which is the case for the relatively cold discs considered in this paper. That version of \galactics\ had a King model for the bulge \citep{King1966}, a lowered Evans model for the halo \citep{evans1993, kuijken1994}, and a disc that is Maxwellian in both $E_z$ and the energy of radial oscillations, $E-E_c$, where $E_c(L_z)$ is the energy of a circular orbit with angular momentum $L_z$ \citep{Kuijken1995}. \citet{Widrow2008} extended the code to allow for more general models of the bulge and halo. One begins with target density profiles and calculates the distribution functions $f_{\rm bulge}(E)$ and $f_{\rm halo}(E)$ via the Eddington inversion formula. Finally, \citet{Deg2019} augmented it to allow for a two-component (e.g. thin$+$thick) stellar disc and a gas disc.

The first step in constructing an equilibrium dynamical model is to calculate the self-consistent gravitational potential from the space densities of the model components.  Since $E$ and $E_z$ are implicit functions of the spatial coordinates, Poisson's equation must be solved iteratively. This is accomplished using an expansion in even Legendre polynomials. A key innovation from \citet{Kuijken1995} was to use an analytic density-potential pair to capture the short-wavelength component of the disc potential. Very accurate density-potential pairs can be calculated with Legendre polynomials up to order $10$.

\subsection{Stellar Discs}
\label{ssec:StellarDiscs}

In previous implementations of \galactics, the mid-plane density and radial velocity dispersion were both assumed to be exponential functions of galactocentric cylindrical radius $R$. 
The exponential disc is motivated by the seminal work of \citet{Freeman1970}, who found that the surface brightness profiles of disc galaxies outside the bulge were well fit by an exponential profile. In the MW, \citet{Bovy2013} found that the surface density profile at $5 \leq R/\kpc \leq 10$ was approximately exponential. 
In addition, the vertical velocity dispersion was tuned to give a vertical scale height that was approximately constant in $R$.  These choices were made to match the qualitative features of edge-on galaxies \citep{Bottema1993} and yielded models where the surface density profile was approximately exponential.

In this work, we consider models with S\'{e}rsic discs, which are straightforward to implement in \galactics. A full description of this implementation is presented in Appendix \ref{sec:Galactics}, and the new version of \galactics\ is publicly available through GitHub (see footnote 1).  Armed with this new version of \galactics\ we can generate bulgeless models.

\subsection{Model Details}\label{ssec:ModelDetails}

Our starting point for the suite of MW models are the \citet{D'Onghia2020} and \citet{TepperGarcia2021} models.  Rather than include a classical bulge, we fit the parameters of a S\'{e}rsic disc to the combined stellar surface density of the bulge + disc systems used in those models.  The \citet{D'Onghia2020} model used the \textsc{AGAMA} software package \citep{Vasiliev2019} to produce a 92.4 million particle simulation.  Their ICs consisted of a Hernquist bulge \citep{Hernquist1990}, an exponential disc, and a Hernquist halo.  The bulge had $M_{B}=8\times10^{9}$ \msol\ and $a_{B}=120$ pc; the disc had $M_{d}=4.8\times10^{10}$ \msol\ and $R_{d}=2.67$ kpc; and the halo had $M_{DM}=10^{12}$ \msol\ and $a=30$ kpc.  The ICs of the \citet{TepperGarcia2021} model were produced using \textsc{AGAMA}.  Their Hernquist bulge had $M_{B}=1.3\times10^{10}$ \msol\ and $a_{B}=0.6$ kpc, their NFW \citep{Navarro1997} dark matter halo had $R_{h}=19$~kpc and $\rho_{0}=9\times10^{9}$ \msol\ kpc$^{-3}$ and their exponential disc had $M_{d}=4.3\times10^{10}$ \msol\ and $R_{d}=2.5$~kpc.

In order to build versions of these models, it is necessary to convert the \citet{D'Onghia2020} and \citet{TepperGarcia2021} parameterizations into equivalent \galactics\ parameterizations. The version of \galactics\ used here consists of a double-power law DM halo with a density given by
\begin{equation}
    \rho_{h}(r)=\frac{2^{1-\alpha}\sigma_{h}^{2}}{4\pi R_{h}^{2}}\frac{1}{(r/R_{h})^{\alpha}(1+r/R_{h})^{\beta-\alpha}}C(R_{h,t},\delta_{R,h,t})~,
\end{equation}
where $\sigma_{h}$ and $R_{h}$ are the scale velocity dispersion and radius respectively, $\alpha$ is the inner slope, $\beta$ is the outer slope, and $C(R_{h,t},\delta_{R,h,t})$ is a truncation function with $R_{h,t}$ and $\delta_{R,h,t}$ being the truncation radius and truncation width respectively.  The S\'{e}rsic disc has a density given by
\begin{equation}
\begin{split}
    \rho_{d}(R,z)=\frac{M_{d}}{4\pi R_{d}^{2}z_{d}\Gamma(2n)}e^{-(R/R_{d})^{1/n}} \\ \times\mathrm{sech}^{2}\left(\frac{z}{z_{d}}\right)C(R_{d,t},\delta_{R,d,t})~,
\end{split}
\end{equation}
where $M_{d}$ is the disc mass, $R_{d}$ is the disc scale length, $z_{d}$ is the scale height, $n$ is the S\'{e}rsic index, and $C(R_{d,t},\delta_{R,d,t})$ is a truncation function for the disc.  The \galactics\ gas disc surface density is given by
\begin{equation}
    \Sigma_{g}(R)=\frac{M_{g}}{2\pi R_{g}^{2}}e^{-R/R_{g}}C(R_{g,t},\delta_{R,g,t})~,
\end{equation}
where $M_{g}$ is the gas mass, $R_{g}$ is the gas scale radius, and $C(R_{g,t},\delta_{R,g,t})$ is a truncation function for the gas disc.  The gas disc is assumed to be in hydrostatic equilibrium at temperature $T_{g}$. The scale height is then a function of R and is set by the condition that the gas pressure balances the gravitational force toward the mid plane.

We start with two fiducial gasless models, one approximating the \citet{D'Onghia2020} model, which we refer to as model D00, and the other approximating the \citet{TepperGarcia2021} model, which we refer to as model TG00. The \galactics\ parameters for these models are listed in Table \ref{Tab:ModelParams}.  We also consider a sequence of models based on D00 with the same halo and stellar disc parameters that also include gas discs with masses equal to 7\%, 15\%, 20\%, and 30\% of the stellar disc mass (termed D07, D15, D20, and D30, respectively).  The gas discs in these models have an exponential scale length of $R_{g} = 6.5$ kpc and a temperature of $10^{4}$ K. Since this scale length is about 2.4 times larger than the scale length of the stellar disc, the gas disc increases the rotation curve by only $<5\%$ for D30, the case with the most massive disc.  Finally, we consider three models (TG07, TG07v2, and TG07v3) with an additional gas disc of 7\% the stellar disc's mass. The latter two differ from TG07 in the way the gas disc is initialized and the way feedback is implemented as discussed below.  While the stellar and DM parameters are broadly the same as the D00 and TG00 models, the full parameters of these gaseous models are also listed in Table \ref{Tab:ModelParams}.  The stellar disc velocity dispersions are set such that the Toomre $Q$ parameter is greater than 1 at all radii.  

\begin{table*}
    \centering
    \begin{tabular}{|c|c|c|c|c|c|c|c|c|}
        \hline 
        Parameter & Unit & D00 & D30 & D20 & D15 & D07 & TG00 & TG07 \\
        \hline 
        \multicolumn{9} {c|}{Halo}\\
        \hline 
        $\sigma_{h}$ & $\kms$ & 550 & 550 & 550 & 550 & 550 & 405 & 405 \\ 
        $R_{h}$ & kpc & 30  & 30 & 30 & 30 & 30 & 19 & 19 \\ 
        $\alpha$ &  & 1 &1 & 1 & 1 & 1 & 1 & 1 \\ 
        $\beta$ &  & 3 & 3 & 3 & 3 & 3 & 3 & 3 \\ 
        \hline 
        \multicolumn{9} {c|}{S\'{e}rsic disc}\\
        \hline 
        $M_{d}$ & $10^{10}$~\msol & 5.2  & 5.2  & 5.2  & 5.2  & 5.2   & 5.5  &  5.5  \\ 
        $R_{d}$ & kpc & 0.91 & 0.91 & 0.91 & 0.91 & 0.91 & 0.43 & 0.43 \\ 
        $n$ &  & 1.5 & 1.5 & 1.5 & 1.5 & 1.5 & 1.75 & 1.75 \\ 
        $z_{d}$ & kpc & 0.25 & 0.25 & 0.25 & 0.25 & 0.25 & 0.25 & 0.25 \\ 
        $\sigma_{r1}$ & $\kms$ & 80 & 90 & 80 & 80 & 80 & 80 & 80 \\ 
        $R_{\sigma 1}$ & kpc & 2.5 &  2.3 & 2.8 & 2.5 & 2.5 & 2.5 & 2.5 \\ 
        $\sigma_{r2}$ & $\kms$ & 28 & 0.0 & 30 & 30 & 28 & 70 & 70 \\ 
        $R_{\sigma 2}$ & kpc & 1.8 & 2.0 & 1.8 & 1.8 & 1.8 & 1.8 & 1.8 \\
        \hline 
        \multicolumn{9} {c|}{Gas disc}\\
        \hline 
        $M_{g}$ & $10^{9}$~\msol & - & 17.2 & 10.2 & 7.8 & 3.65  & - & 3.85 \\ 
        $R_{g}$ & kpc  & - & 6.5 & 6.5 & 6.5 & 6.5  & - & 6.5 \\ 
        $T_{g}$ & K  & - & 10$^{4}$ & $10^{4}$ & $10^{4}$ & $10^{4}$ & - & 10$^{4}$ \\ 
        \hline 
    \end{tabular}
    \caption{Parameters for all the models in this paper. All three versions of the TG07 model correspond to the same physical system.
    }
    \label{Tab:ModelParams}
\end{table*}

\subsection{Simulation Details}\label{ssec:SimDetails}

We run the collisionless simulations (D00 and TG00) with {\sc pkdgrav2} \citep{pkdgrav}, a treecode for $N$-body simulations. All the models are initialized with $5\times10^{6}$ dark matter particles and $4.8\times 10^{6}$ stellar particles.    We use a particle softening\footnote{We report the softening spline mid-point as the softening length.} of $\epsilon = 50~ \pc$ for the stars, and $\epsilon = 100~ \pc$ for dark matter particles. We select a base time step $\Delta t = 5~ \Myr$, with timesteps of individual particles refined such that each satisfies the condition $\delta t = \Delta t/2^n < \eta \sqrt{\epsilon/a_g}$, where $a_g$ is the acceleration at the particle’s current position. This results in 7 rungs (i.e. $n=6$, corresponding to a minimum $\delta t = 78,125~ \yr$) in both D00 and TG00. We set $\eta = 0.2$ and the opening angle of the treecode gravity calculation $\theta = 0.7$. We evolve these models for 10~Gyr.

Since {\sc pkdgrav2} is a pure N-body code, and cannot model gas, we run the simulations with gas using the efficient $N$-body$+$SPH code {\sc ChaNGa} \citep{Jetley2008,Jetley2010,Menon2015}, which is a {\sc Charm++} extension of {\sc Gasoline} \citep{wadsley+04, wadsley+17}\footnote{{\sc Gasoline} is itself a hydrodynamics extension of {\sc pkdgrav}.}. The gas disc simulations have $3\times 10^{5}$ gas particles, regardless of the gas mass fraction.
As with the collisionless simulations, we employ a tree opening angle of $\theta = 0.7$ with a base time-step of $\Delta t = 5~ \Myr$. Time-steps of individual particles are then refined in the same way as for the collisionless simulations with two differences: we set $\eta = 0.175$, and the time-steps of gas particles must also satisfy the additional condition $\delta t_{gas} < \eta_{courant}h/[(1 + \alpha)c + \beta \mu_{max}]$, where $\eta_{courant} = 0.4$, $h$ is the SPH smoothing length set over the nearest 32 particles, $\alpha = 1$ is the shear coefficient, $\beta = 2$ is the viscosity coefficient, $c$ is the sound speed and $\mu_{max}$ is the maximum viscous force between gas particles \citep{wadsley+04}. The softening of gas particles, which is inherited by star particles formed from them, is $\epsilon = 50~\pc$.
With these time stepping recipes, 7-9 rungs (maximum $n = 6-8$, corresponding to $\delta t = 78,125-19,531$~yr) are required to move all the particles.

In the D series of simulations star formation requires a gas particle to have cooled below 15,000 K and exceeded a density of $0.1$ amu\,cm$^{-3}$.  Gas particles meeting these criteria form stars with a probability of $0.05$ per dynamical time \citep{Stinson2009}, i.e. the star formation efficiency is set to 5\%.  Chemical and thermal mixing use the prescriptions of \citet{shenwadsley2010}. In models D07 and D15 the supernova feedback couples $0.8 \times 10^{51}$ erg per supernova to the gas via the superbubble prescription of \citet{keller2014}. In model D30 we dial down the strength of the supernova feedback to $0.4 \times 10^{51}$ erg per supernova. In all these models stars form with a mass $1.1 \times 10^4$~\msol\ and gas particles are removed and their remaining mass distributed to the neighbouring gas particles when their mass drops below $1.1~\times~10^4$~\msol.

The TG series of models with gas all have an additional $7\%$ of the stellar disc mass in gas. They are run with the same base time-step, time step refinement parameters, star formation efficiency, star formation density threshold, and supernova feedback strength ($0.8 \times 10^{51}$ erg per supernova). Stars form with an initial mass $1.1 \times 10^4$~\msol\ in models TG07 and TG07v2, while in TG07v3 we reduce this to $4.2~\times~10^3$~\msol.


\section{Evolution of S\'ersic disc models}
\label{Sec:GasAndBars}

\subsection{The D-series of simulations}

\begin{figure*}
\centering
    \includegraphics[width=\textwidth]{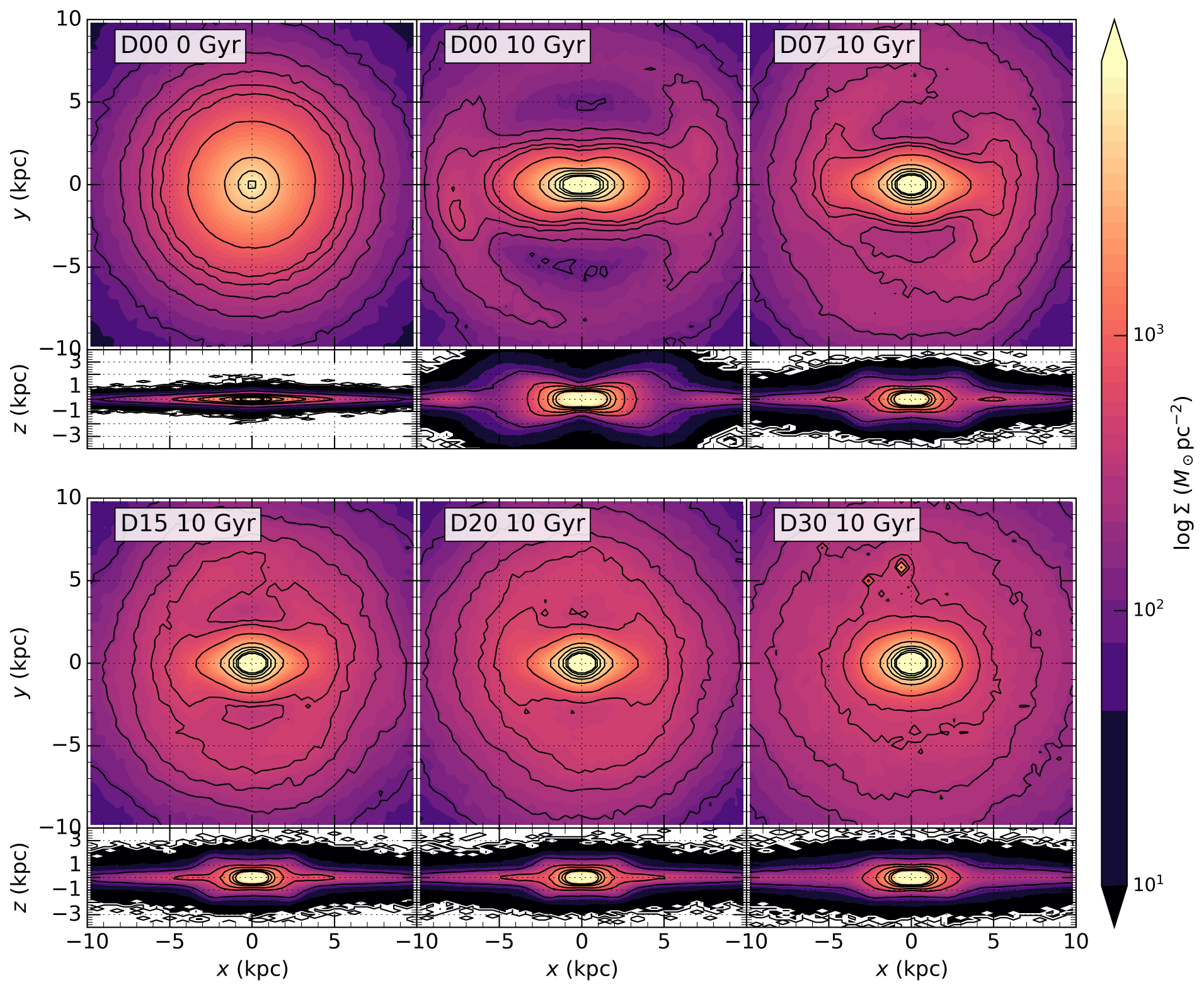} 
    \caption{A comparison of the inner stellar disc for the D sequence models.  The upper left panel pair shows the surface density and cross-section of the stellar disc for each D model at $T=0$ Gyr, while the other panels show the stellar disc of the D-series at $t=10$ Gyr.  In all 10 Gyr panels, the stellar discs are rotated to place the bar along the $x$-axis. For the $(x,z)$-plane views, we have imposed a cut $|y|<1$ kpc to emphasize the BP-shaped nature of the bulge.
    }
  \label{Fig:DOnghiaStellarSDMaps}
\end{figure*}

We start by considering the D-series of models, in which we vary the gas fraction from $0\%$ to $30\%$. In Figure \ref{Fig:DOnghiaStellarSDMaps} we compare face-on and cross-sectional views of the projected stellar density for the different D-series models (the cross-sections highlight the BP bulge shape). The upper left panel shows the initial disc while the other panels show the discs at $t=10$ Gyr.  As expected, the longest bar, which reaches 8 kpc, is seen in the gasless (D00) simulation, highlighting the problem of runaway secular bar growth. It also has the most prominent BP bulge.
The strength of the bar and of the BP-bulge visibly decrease with increasing gas fraction. Even with a modest 7\% gas disc (D07), the final bar and BP~bulge size is decreased by a factor $\sim 2$ over model D00.

\begin{figure*}
\centering
    \includegraphics[width=0.8\textwidth]{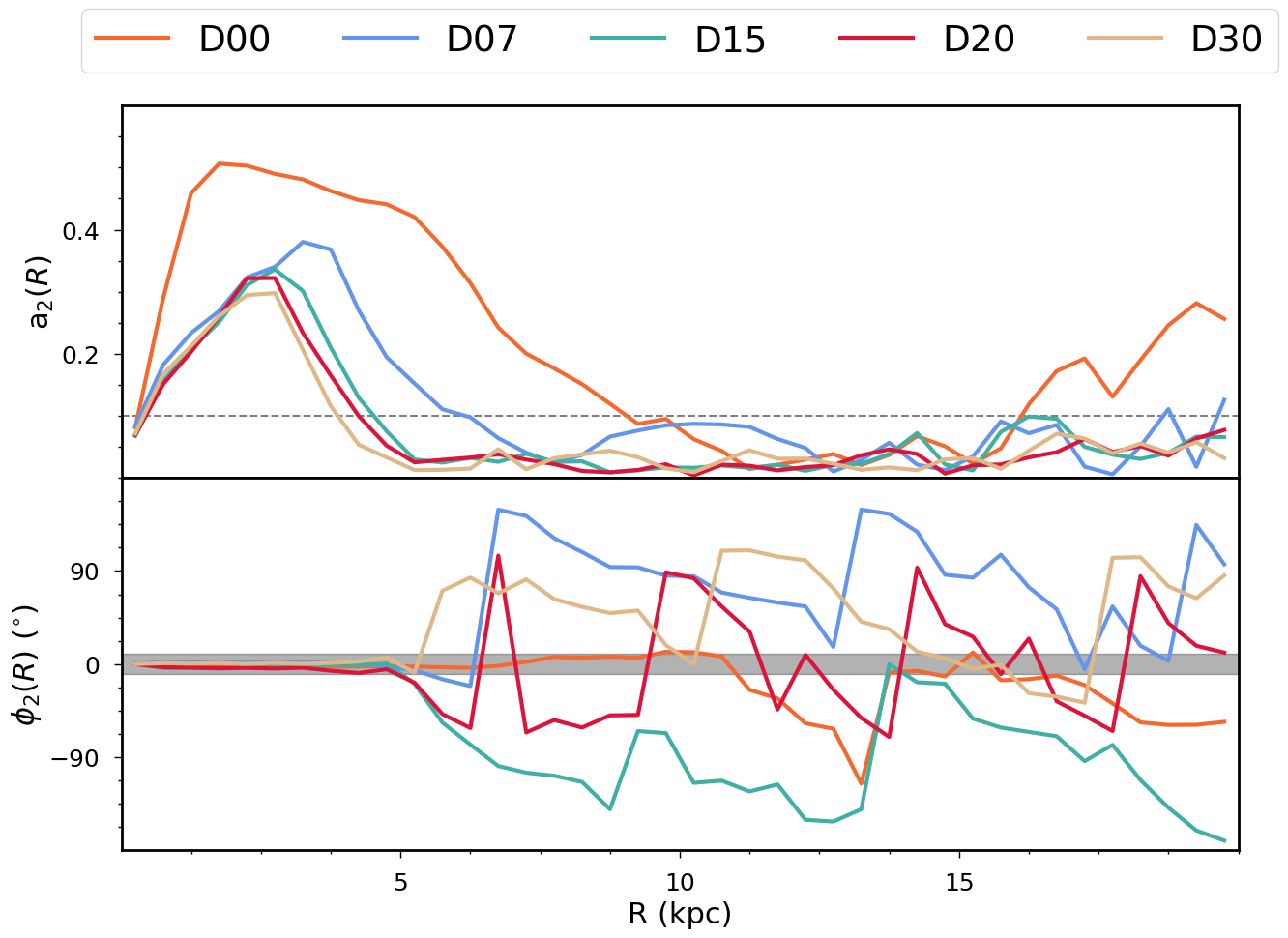} 
    \caption{A comparison of the 2nd Fourier moment amplitude (top panel) and phase (bottom panel) for the D-sequence of models at $t=10~\textrm{Gyr}$.  The bar angle has been set relative to the value of $\phi_2$ in the innermost radial bin.  The dashed line in the upper panel shows the $a_{2}<0.1$ limit, while the grey shaded region in the bottom panel shows the size of the $|\delta \phi_2| <10^{\circ}$ which are both used to determine the bar length in Sec. \ref{ssec:DOnghiaBarEvolve}. In this plot, the bin size is 0.5 kpc.}
  \label{Fig:DOnghiaBarProfiles}
\end{figure*}

The bar strength can be quantified by the amplitude of the $m=2$ Fourier moment:
\begin{equation}
    a_{2}(R) = \left| \frac{\sum_k m_k e^{2 i \theta_k}} {\sum_k m_k}\right|,
\end{equation}
where the sum is over a cylindrical ring of radius $R$, and $m_{k}$ and $\theta_{k}$ are the mass and angle of the $k^{\rm th}$ particle. The phase angle of the $m=2$ Fourier mode is
\begin{equation}
    \phi_2(R)=\frac{1}{2}\textrm{tan}^{-1}\left(\frac{\sum_k m_k \textrm{sin}(2\theta_k)}{\sum_k m_k \textrm{cos}(2\theta_k)} \right)~,
\end{equation}
where the summation is again over all particles in some cylindrical ring.  Figure \ref{Fig:DOnghiaBarProfiles} shows the profile of $a_{2}(R)$ for the stellar discs of D-sequence models at $t=10~\textrm{Gyr}$, as well as the angle, $\phi_2$ of the mode relative to the angle in the innermost radial bin. This plot shows that the bar strength and bar length decreases with increasing gas fraction.  It is worth noting that the jumps in the outer radii are partially due to the cyclical nature of $\phi_{2}$ over a range of $180^{\circ}$.  Additionally, once the $a_{2}$ moment is low, the shape is approximately circular, leading to $\phi_{2}$ becoming essentially meaningless.

\subsubsection{Bar evolution}
\label{ssec:DOnghiaBarEvolve}

Fourier profiles such as those in Figure \ref{Fig:DOnghiaBarProfiles} can be used to quantify the bar's total strength and length, and thus study its evolution.  For this work, we set the bar strength, $A_{2}$, to be the maximum of $a_{2}(R)$, using 40 radial bins with widths of 0.5 kpc.  There are a variety of different methods for calculating the bar length (for examples, see \citealt{Agurerri2000,Athanassoula2002,erwin2005,Michel-Dansac2006,Anderson2022}). Here we follow a similar (but not exactly the same) approach as \citet{Anderson2022}, and calculate both the radius where $a_2(R)$ drops below 0.1 (the grey dashed line in the upper panel of Figure \ref{Fig:DOnghiaBarProfiles}) and where $\phi_2(R)$ changes by more than $10^\circ$ (the grey shaded region in the bottom row of Figure \ref{Fig:DOnghiaBarProfiles}).  The bar length is set as the average of these two radii, while the uncertainty is based on the sum of half the difference between the two radii and half the bin size added in quadrature.

\begin{figure*}
\centering
    \includegraphics[width=0.8\textwidth]{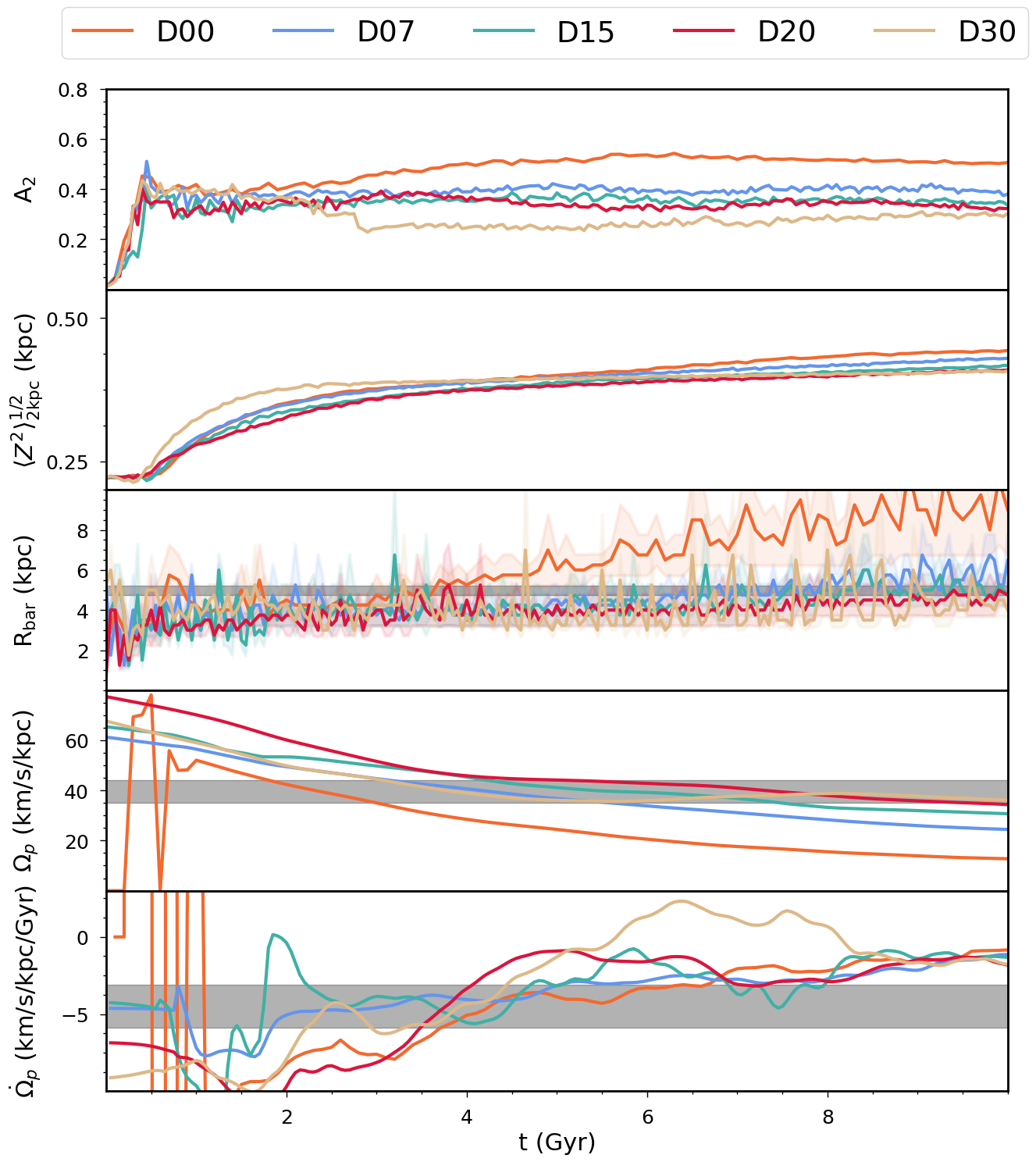} \caption{The evolution of the bars in the `D'-series of models.  From top to bottom the panels show the maximum of $A_{2}(R)$ as the bar strength, the rms of the disc vertical height within the inner 2 kpc, the bar length, the bar pattern speed, and the bar slowdown rate.  The dark grey shaded regions in the bar length, pattern speed, and slow down rate panels are measurements from \citet{Wegg2015}, \citet{Portail2017}, and \citet{Chiba2021} respectively.  The other colored shaded regions in the bar length panel are the uncertainties in the bar length. }
  \label{Fig:DOnghia_Bar}
\end{figure*}

Figure \ref{Fig:DOnghia_Bar} shows the global bar strength, length, and other bar properties as a function of time for the D sequence of models.  Table \ref{tab:BarProperties} lists the bar strength, inner thickness, length, pattern speed, and slow down rate along with the MW values (for those with measured values) at the best matching snapshots.  In all of the D sequence simulations, $A_2$ rises to an initial value between $0.3$ and $0.5$ within the first $500\,{\rm Myr}$, as the bar instability sets in almost immediately and saturates after a few dynamical times. The trend of bar strength decreasing with increased gas fraction in Figure \ref{Fig:DOnghia_Bar} matches the instantaneous $a_{2}(R)$ profiles seen in Figure \ref{Fig:DOnghiaBarProfiles}.

\begin{table*}
    \centering
    \begin{tabular}{|c|c|c|c|c|c|}
        \hline
        Snapshot & $A_{2}$ & $<Z^{2}>^{1/2}$ & $R_{\rm{bar}}$ & $\Omega_{p}$ & $\dot{\Omega}_{p}$  \\
                & & kpc & kpc & km/s/kpc & km/s/kpc/Gyr \\
        MW        & & & $4.6\pm0.3$ &$39.0\pm 3.5$ & $-4.5\pm~1.4$\\
        \hline 
        D00 - 3 Gyr & 0.45 & 0.38 & $4.75\pm1.0$ & 35.& -7.6 \\
        D07 - 5 Gyr & 0.42 & 0.39 & $4.5\pm0.8$ & 37.& -3.1\\
        D15 - 6 Gyr & 0.37 & 0.39 & $4.\pm0.8$ & 39. & -1.4\\
        D20 - 6 Gyr & 0.33 & 0.39 & $4.\pm0.8$ & 42. & -1.6\\
        D30 - 5 Gyr & 0.25 & 0.39 & $3.5\pm0.4$ & 36. & -1.3\\
        \hline
        TG00 - 2.5 Gyr & 0.45 & 0.38 & $5.\pm0.8$ & 33.& -6.3\\
        TG07 - 3 Gyr   & 0.40 & 0.37 & $4.25\pm0.5$ & 44. & -8.\\
        TG07v2 - 3 Gyr & 0.37 & 0.37 & $4.25\pm0.5$ & 45.& -6.7\\
        TG07v3 - 3 Gyr & 0.37 & 0.37 & $4.25\pm1.0$ & 42.& -5.7\\
        \hline 
    \end{tabular}
    \caption{The measured bar properties of the MW, the D simulations, and the TG simulations.  For the simulations, the specific snapshot is selected based on the quality of the combined comparison to the three MW measurements.}
    \label{tab:BarProperties}
\end{table*}

The second row of Figure \ref{Fig:DOnghia_Bar} shows the root mean square thickness of the inner ($R<2\kpc$) disc, $\langle z^2\rangle^{1/2}$. This thickness steadily increases over the course of the simulation reaching a final thickness of about $1.2\,{\rm kpc}$ or a factor of $4-5$ times the initial thickness in the absence of gas, and by a factor of $2-3$ when gas is present. None of the simulations show the sharp increase in thickness that is characteristic of a buckling event. We also computed the \textrm{buckling} amplitude $A_{\textrm{buck}}$ \citep{Debattista2006} and did not find any evidence of major buckling events. We conclude that the bars and BP-bulges formed in the D sequence are formed via resonant trapping rather than a buckling event, consistent with the results of  \citet{SellwoodGerhard202x}, who find buckling is suppressed when the centre of a galaxy is dense.

In the third row of Figure \ref{Fig:DOnghia_Bar}, we plot the bar radius, $R_{\rm bar}$. While all bars start out with the same size ($R_{\rm bar} \simeq 3~\kpc$) as expected, the bar grows longest in the D00 model, reaching $R_{\rm{bar}}>8~\rm{kpc}$ at $t=10~\rm{Gyr}$.  The next longest final bar is found in the D07 model, which reaches $R_{\rm{bar}}\simeq 6 ~ {\rm kpc}$. The simulations with larger gas fractions end with bars having $R_{\rm bar}\simeq 4-5\,{\rm kpc}$, consistent with the observational results of \citet{Wegg2015} who measured, for the MW, $R_{\rm bar} = 4.6\pm0.3\,\rm{kpc}$  from red clump giant stars.

We show the pattern speed, $\Omega_p = d\phi_2/dt$, in the fourth row of Figure \ref{Fig:DOnghia_Bar}, along with the MW measurement of \citet{Portail2017}. The pattern speed is obtained using the single snapshot method of \citet{Dehnen+2023}.  After calculating the pattern speed, we apply the \textsc{lowess} smoothing algorithm from \textsc{SciPy} \citep{2020SciPy-NMeth} to the calculated values to smooth out the numerical fluctuations. 
Unsurprisingly, the bar in model D00 has the slowest final pattern speed. The shorter bars in the simulations with gaseous discs have higher pattern speeds. The bar in each simulation slows down as they grow in strength, and length. The pattern speeds in all the simulations with gas discs are consistent throughout much of their evolution with the MW measurement of $\Omega_{p} = 39.0\pm 3.5~\kms~{\rm kpc}^{-1}$ of \citet{Portail2017}.
Except for model D30, the bars in the models slow down uniformly. Model D30 experiences a period of acceleration during $t \simeq 6-8$ Gyr, during which the bar weakens slightly, before then slowing down again, and reaching values of $\dot{\Omega}_{p}$ very similar to that of the other models, despite having the weakest bar.

By modelling the Hercules Stream as resulting from stars trapped at the bar's corotation resonance, \citet{Chiba2021} presented evidence that the pattern speed of the MW's bar is declining at 
$\dot{\Omega}_{p} = -4.5\pm~1.4~\kms~\rm{kpc}^{-1}~\rm{Gyr}^{-1}$.
We present the \citet{Chiba2021} measurement and $\dot{\Omega}_p$ for the models in the fifth row of Figure \ref{Fig:DOnghia_Bar}.  We calculate the slowdown rate of our simulations by taking the derivative of the smoothed pattern speed.  Since the pattern speed is not perfectly smooth, the calculated $\dot{\Omega}_{p}$ shows a great deal of variation. Unlike the other bar properties presented in Figure \ref{Fig:DOnghia_Bar}, $\dot{\Omega}_{p}$ is not directly correlated with the gas fraction. 
Compared to the \citet{Chiba2021} estimate, the slowdown rates of the D-sequence of models are lower (i.e. closer to zero) for the majority of their evolution.  It is worth noting that, by definition, no simulation can satisfy both the \citet{Portail2017} pattern speed measurement and the \citet{Chiba2021} slow down rate of the MW for longer than $\sim1~\rm{Gyr}$. 
In the case of the gaseous D-sequence of models, the pattern speed and slowdown rate broadly agree with both \citet{Portail2017} and \citet{Chiba2021} around $t\simeq 4-5$~Gyr, but at those times, the bar length's are smaller than the MW's bar.

\subsubsection{BP bulges}

\begin{figure*}
\centering
    \includegraphics[width=0.8\textwidth]{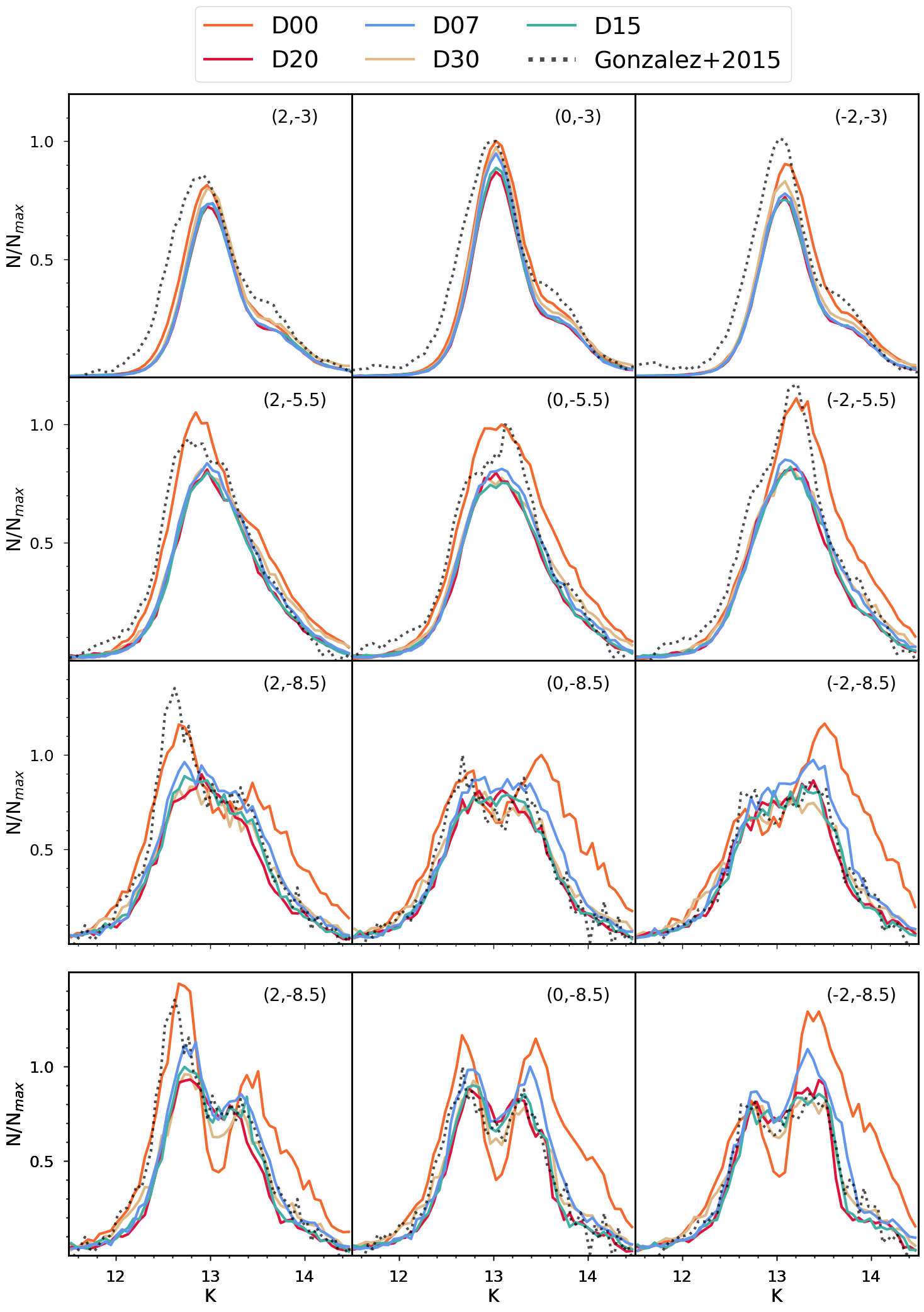} 
    \caption{Mock observations of the simulations along different lines-of-sight towards the BP-bulge for the D sequence of models at $t=10~\rm{Gyr}$.  The upper three rows have all been convolved with the observed widths of both red clump and red giant branch bump stellar magnitudes, while the bottom row is unconvolved.  In other words, the bottom row shows the underlying magnitude distribution of stars along the $b=-8.5^{\circ}$ lines-of-sight, while the third row shows that same distribution convolved by the observational distribution of red clump and red giant bump branch stars.  The model curves in the top three rows are normalized to the peak value of all model curves in the $l=0^{\circ}$ panels in each row. The bottom row model curves are normalized using the exact same factor as the third row in order to highlight the effect of the convolution on the underlying particle distribution.  The \citet{Gonzalez2015} data have been normalized separately to the peak in the $l=0^{\circ}$ panels of each row. The galactocentric coordinates of the lines-of-sight are listed in the upper right corners of each panel.
    }
  \label{Fig:DOnghiaBulgeLOS}
\end{figure*}

In the MW, the presence of a BP bulge means that, along certain lines of sight, two peaks in the number counts of stars as a function of distance are evident \citep{McWilliamZoccali2010, Nataf+2010, Saito+2011, Wegg2013, Gonzalez2015}.  In Figure \ref{Fig:DOnghiaBulgeLOS} we present mock observations of the number density of simulation particles along varying lines of sight together with the observations of \citet{Gonzalez2015}. The observations include both red clump and red giant branch bump (RGBB) stars.  To generate the model curves, all particles within a projected $1^{\circ}$ of the line-of-sight are randomly assigned the absolute magnitude of a red clump or a RGBB star.  In the upper three rows the red clump absolute magnitudes are drawn from a Gaussian with $\left<M\right>=-1.55$ and $\sigma_{\rm{RC}}=0.17$, while the RGBB magnitudes are drawn from a Gaussian with $\left<M\right>=-0.84$ and a $\sigma_{RGBB}=0.17$ \citep{Nataf2013,Gonzalez2015,Gonzalez2018}.  In the bottom row, all particles are given the absolute magnitude of either a red clump or RGBB star.  In all cases, the ratio of red clump to RGBB stars is set to $20\%$ \citep{Wegg2013}.  Once a particle is assigned an absolute magnitude, its apparent magnitude is calculated using the distance modulus, and then binned to create the histograms seen in Figure \ref{Fig:DOnghiaBulgeLOS}.  As such, the bottom row shows the specific snapshot's true apparent magnitude (or distance) distribution along the $l=-8.5^{\circ}$ lines-of-sight rather than the convolved distribution. The model histograms in the upper three rows are normalized by the singular peak of all the models in the $l=0^{\circ}$ panel of each row in order to highlight the differences between the models.  The fourth row (i.e the unconvolved magnitude distributions) are instead normalized by the same factor as the third row so that the effect of the convolution by the distribution of stellar magnitudes can be seen clearly.
The \citet{Gonzalez2015} data is normalized separately by its own peak in the $l=0^{\circ}$ panels (matching the normalization of Figure 2 of \citealt{Gonzalez2015}). \

When comparing the bulge lines-of-sight to the \citet{Gonzalez2015} data it is important to compare the shapes rather than the height of the curves due to our use of a single normalization for all models.  This normalization allows a comparison between the different models with regards to how they evolve. But for direct comparisons to the \citet{Gonzalez2015} data, individual normalizations are preferred (see Appendix \ref{app:supplemental} for an example).  Thus, for this discussion, we focus on the shape of the D-family of models and the \citet{Gonzalez2015} data.  In the top row, the most noticeable feature is the secondary peak in the brightness distributions at $\rm{K} \simeq 13.5$ in all panels.  This feature is due to the RGBB stars and not due to the BP bulge structure. 
 Comparing the models to the \citet{Gonzalez2015} data, it is clear that the observations have a broader distribution as well as a larger secondary peak.  The extra broadening is likely due to remaining differential reddening in the data \citep{Gonzalez2018}. The difference in the secondary peak sizes may be due to the ratio of RC to RGBB stars along these lines-of-sight.  The key result of the $b=-3^{\circ}$ panels is that the primary peaks are located at the same magnitudes in the models and the data.

The $b=-5.5^{\circ}$ panels are more interesting.  The model profiles are broadly flat compared to the data, which shows asymmetrical peaks at $l=2^{\circ}$ and $l=-2^{\circ}$.  However, all the models that include gas have similar width as the \citet{Gonzalez2015} data, while the D00 model is broader.  This is due to the extremely large box-peanut (BP) bulge present in the D00 model, while the rest have much more MW-like sizes.  This result is made more clear in the two $b=-8.5^{\circ}$ rows.  The unconvolved profiles (bottom row) all show strong bimodalities, but, when convolved with the appropriate stellar distribution widths, much of this bimodality disappears (third row).  Nonetheless, the models that include a gas disc do show differences between the near side ($\rm{K}\sim 12.6$) and far side ($\rm{K}\sim 13.4$) peaks that are similar to the \citet{Gonzalez2015} data and are caused by their BP bulge structure.  The distributions have similar widths, the peaks are in the correct locations, and, for the $l=2^{\circ}$ panel, there are more nearby stars than distant stars, which is flipped for the $l=-2^{\circ}$ panel. On the other hand, the D00 model is much broader than the observed data, with peaks at the incorrect locations.   Ultimately, the D07 model has the most similar shape to the \citet{Gonzalez2015} data across all panels (see further discussion in Appendix \ref{app:supplemental}), 
but the other models gas disc models are also reasonable.  Nonetheless, given the disagreements with the bar parameters, we conclude that none of the D-series models are a close match to the MW.

\subsection{A more realistic MW bar model}\label{SSec:TG07Sims}

The models in the D-series show that reasonably-sized bars {\it can form} in a S\'ersic disc, and that the presence of gas can reduce the secular growth rate of the bar for a final model that is not too different from the MW. In order to produce an improved model of the MW, we turn to the models based on the \citet{TepperGarcia2021} model.  The \citet{TepperGarcia2021} model was designed to match multiple observables of the MW, including the Galactic rotation curve (RC) and surface density (SD) profiles. Although the \citet{TepperGarcia2021} model evolves away from these initial constraints, it provides a better-tuned starting point.  In addition, it is more disc dominated in the inner region and therefore more susceptible to the bar instability, while potentially lowering the secondary secular growth.

Based on the results of the D-sequence models, we only consider a gasless model, TG00, and three instances with a 7\% gas disc, (TG07, TG07v2, and TG07v3).  TG07v2 is designed to examine the effect that bar stochasticity \citep{sellwood_debattista09}, while TG07v3 is designed to investigate the effects of different star formation subgrid parameters (see Section~\ref{Sec:ICs}).

\subsubsection{Bar evolution}

\begin{figure*}
\centering
    \includegraphics[width=0.8\textwidth]{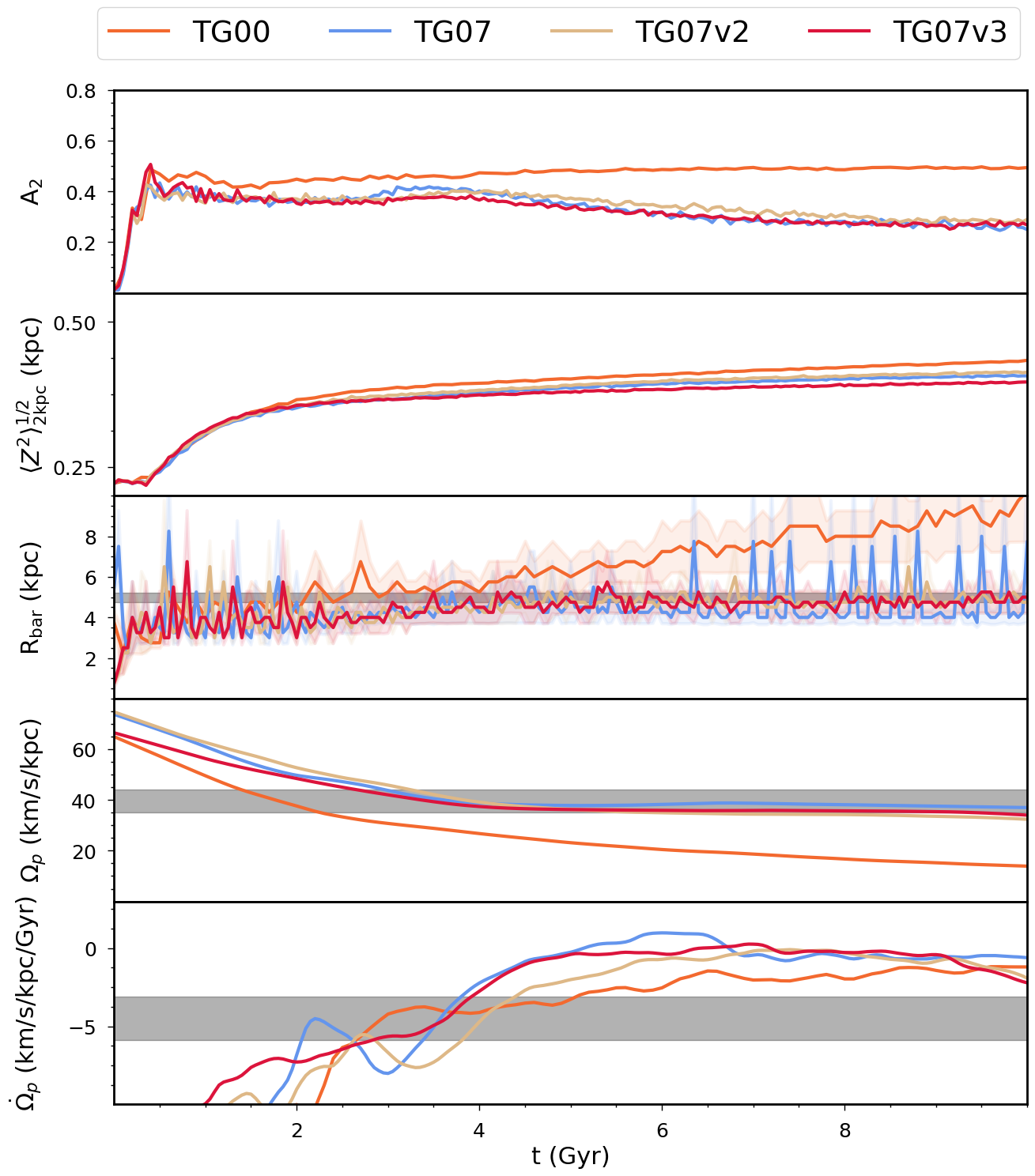} \caption{The bar evolution of the TG-sequence of models as function of time.  From top to bottom the panels show the maximum of $a_{2}(R)$ as the bar strength, the rms of the disc vertical height within the inner 2 kpc, the bar length, the bar pattern speed, and the bar slowdown rate.  The grey shaded regions in the bar length, pattern speed, and slowdown rate panels are the same as in Figure~\ref{Fig:DOnghia_Bar}, while the other coloured shaded regions in the bar length panel are the uncertainties in the bar length.}
  \label{Fig:TG07_Bar}
\end{figure*}

Figure \ref{Fig:TG07_Bar} shows the time evolution of bar properties for the TG-sequence of models using the same analysis methods as in Figure \ref{Fig:DOnghia_Bar}.  Like the D-sequence, all TG simulations rapidly develop a bar which initially extends to $R_{\rm{bar}}\simeq 4$~kpc.  In model T00 the bar becomes far stronger and more extended than in the MW, reaching $R_{bar}\sim~10$~kpc, and a pattern speed of $\sim 20~\kms~\rm{kpc}^{-1}$, which is comparable to the evolution seen in model D00. 

The models with gas discs evolve differently than the equivalent D07 model, as all three of the TG07-sequence reach a steady configuration by $\sim~3.5-4~\rm{Gyr}$, with little evolution in $R_{bar}$ or $\Omega_p$ thereafter.  There is a slight weakening of the bar strength, $A_{2}$, over this period.  The fact that all three $7\%$-gas models develop bars with constant lengths and pattern speeds despite the different random initializations and star formation recipes suggests that this stability is numerically robust.  It is likely that these are in a metastable configuration (see Appendix \ref{app:metastability} for a more detailed discussion).  TG07 briefly attains a positive torque, i.e. $\Omega_p$ increases, over a period of $\sim 1$ Gyr, which may be associated with the fact that this model briefly has a stronger bar than any of the other 7\% gas disc models. The thickness of the TG models evolves very slowly, with no evidence of buckling in any of them.

\subsubsection{BP bulge}

Given the similarity of the three 7\% gas disc TG models, we consider the BP bulge only in the TG07 model. The same analysis for models TG07v2 and TG07v3 is presented in Appendix~\ref{app:supplemental}.

\begin{figure*}
\centering
    \includegraphics[width=\textwidth]{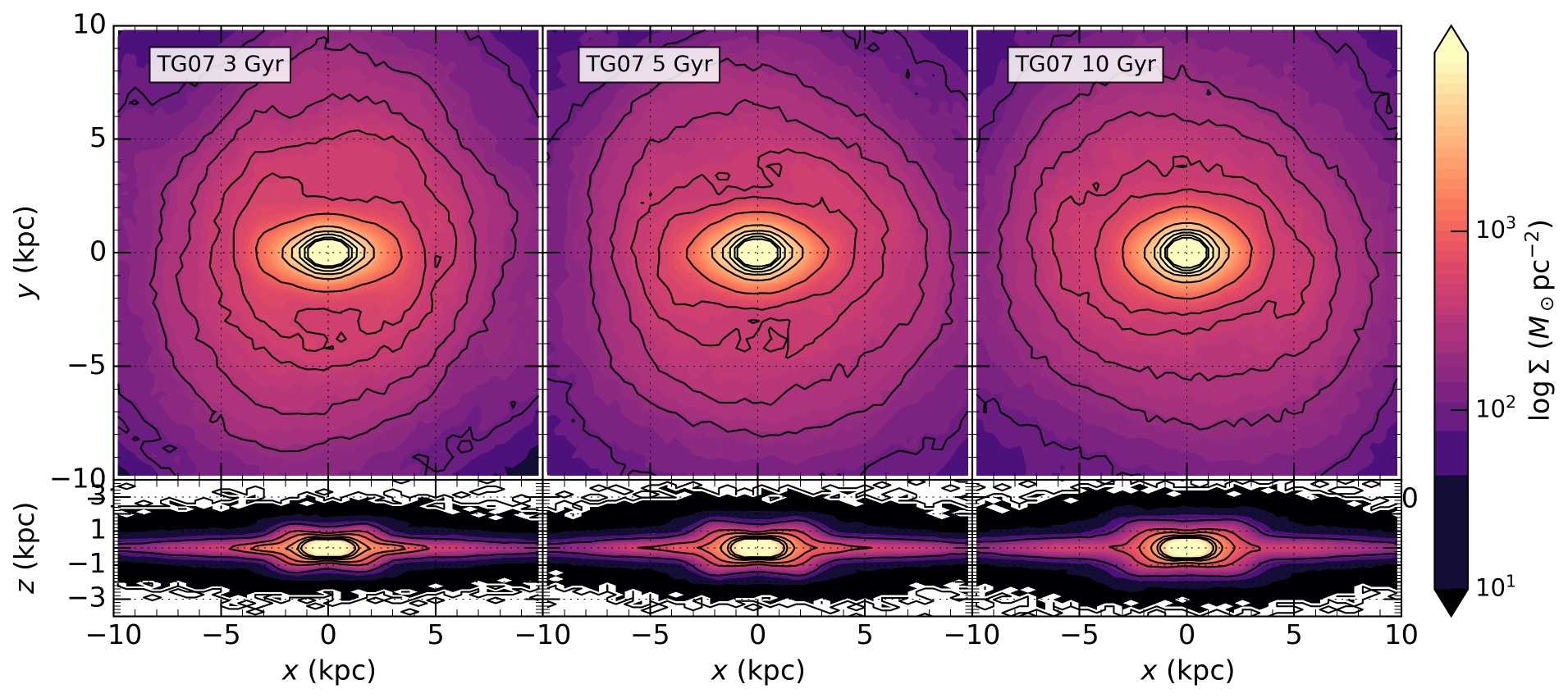} 
    \caption{A comparison of the surface density maps and 2 kpc-wide cross-sections of the stellar disc of the TG07 model at different times.  In all panels, the stellar discs are rotated to place the bar along the $x$-axis. For the $(x,z)$-plane views, we have imposed a cut $|y|<1$ kpc to emphasize the X-shaped nature of the bulge.}
  \label{Fig:TepperGarcia_SDMaps}
\end{figure*}

\begin{figure*}
\centering
    \includegraphics[width=0.8\textwidth]{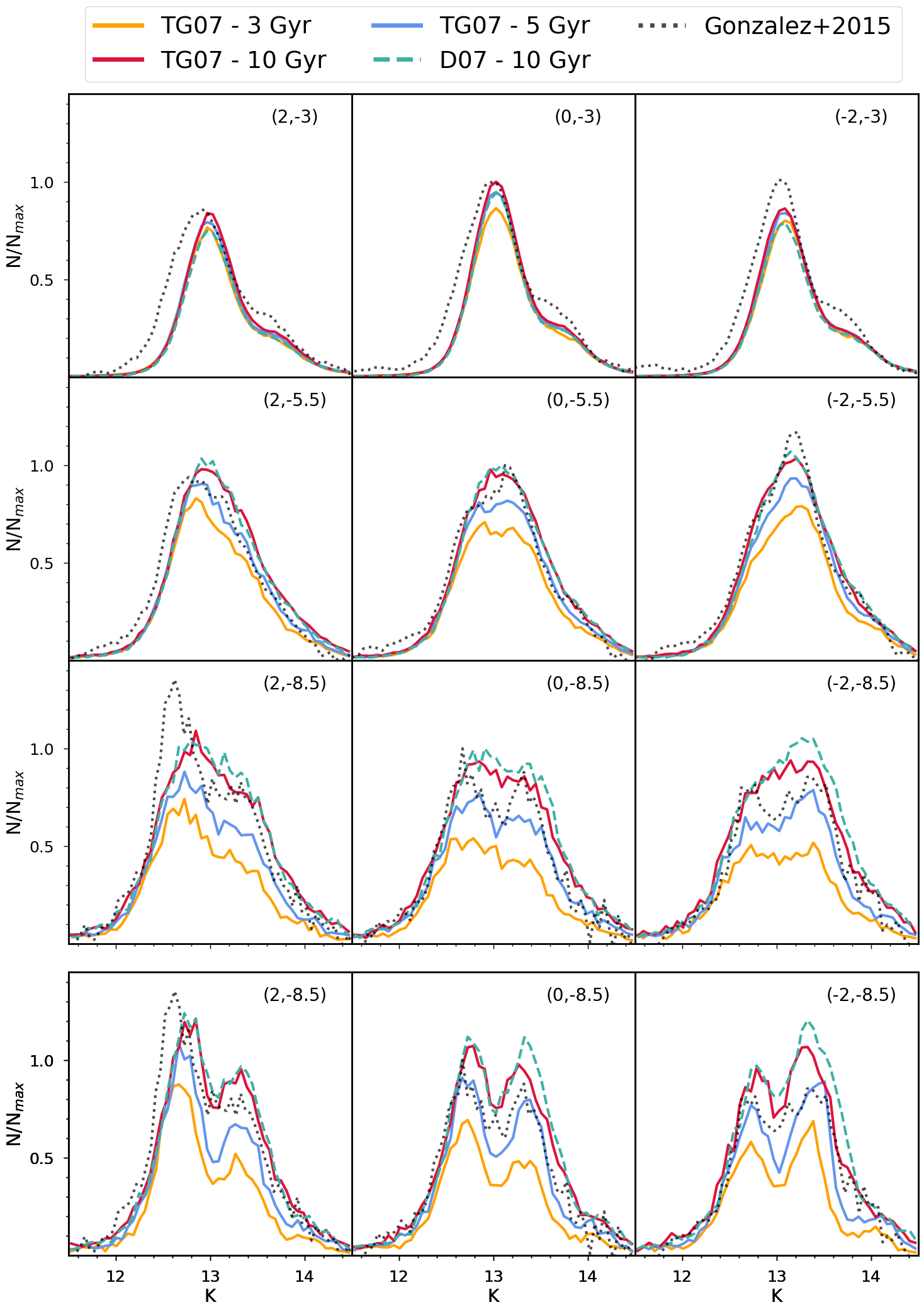} \caption{Mock observations of red clump star and RGBB magnitudes along different lines-of-sight towards the BP-bulge for the TG sequence of models at $t=10~\rm{Gyr}$.  As with Figure \ref{Fig:DOnghiaBulgeLOS}, the model curves in the upper three rows are convolved with the red clump and RGBB magnitudes and normalized by the peak of all model curves in the $l=0^{\circ}$ panels in each row.  The bottom row shows the unconvolved magnitude distribution of the particles and is normalized by the same factor as the third row in order to highlight the effect of the convolution on the particle distribution.  The \citet{Gonzalez2015} data have been normalized separately to the peak in the $l=0^{\circ}$ panels of each row.   The galactocentric coordinates of the lines-of-sight are listed in the upper right corners of each panel.
    }
  \label{Fig:TepperGarcia_BulgeLOS}
\end{figure*}

Figure \ref{Fig:TepperGarcia_SDMaps} shows the surface density map and cross-section of model TG07 at 3, 5, and 10 Gyrs.  Some weak evolution is evident, with the bar becoming slightly rounder over time, which is consistent with Figure~\ref{Fig:TG07_Bar}, which shows that the bar weakens somewhat.  Additionally, the BP bulge becomes slightly thicker with a less peanut/more boxy shape developing. This can be seen more clearly in Figure~\ref{Fig:TepperGarcia_BulgeLOS}, which shows the mock red clump bimodality at $t=3$, $5$, and $10~\rm{Gyr}$ compared with the observational data from \citet{Gonzalez2015}.  Once again, it is important to note that, for Figure \ref{Fig:TepperGarcia_BulgeLOS}, the normalization to a single snapshot peak rather than individual normalizations enables an examination of the time evolution of the BP shape. As with Figure \ref{Fig:DOnghiaBulgeLOS}, in Figure \ref{Fig:TepperGarcia_BulgeLOS}, we compare the shape of the curves in each panel to the \citet{Gonzalez2015} data rather than their heights (see Appendix \ref{app:supplemental} for further discussion).
At $3$ Gyr the bar is still evolving, but has nearly reached the MW's bar length and pattern speed. At this point, the differences between the low and high magnitude regions are the largest, and the overall distribution across all lines-of-sight is most similar to the \citet{Gonzalez2015} data.  As the bar and BP bulge broaden laterally, the line-of-sight distributions become smoother and the difference between the near and far sides of the stellar distributions decrease.  Thus the TG07 model at $t=3~\rm{Gyr}$ is the closest to matching the MW. 

While the TG07 model at $t=3~\rm{Gyr}$ is the closest to matching the MW, the central dips in the $l=-8.5^{\circ}$ panels are missing, and there are slight differences in the location of the peaks and ratio of the approaching/receding side number counts.  We attribute these to the TG07 model having a weaker BP signature than the actual Galaxy. To test this, we explored adjustments to the ratio of red clump/RGBB stars to mimic possible uncertainties in this ratio.  Such adjustments only change the height and slope of the inflection point seen in the $l=-3^{\circ}$ panels.  We also explored a range of intrinsic widths to the stellar magnitudes of the red clump stars.  While a smaller width can lead to the intrinsic bimodality seen in the bottom row of Figure \ref{Fig:TepperGarcia_BulgeLOS} being observed in the 3rd row, it also shrinks the full distribution width in the top row as well as adjusting the ratio of approaching/receding side number counts.  Given that the intrinsic width is based on the observations of \citet{Nataf2013} and \citet{Gonzalez2015} and that the fits in Figure \ref{Fig:TepperGarcia_BulgeLOS} are superior to any tested alternatives, we are left to conclude that the remaining differences between the \citet{Gonzalez2015} observations and the TG07 model are truly due to differences in the BP structure.  This result highlights the fact that, while this model does reasonably well at producing the main features of the bulge, more work will be required for a detailed match to the MW.

\begin{figure}
\includegraphics[angle=0.,width=\hsize]{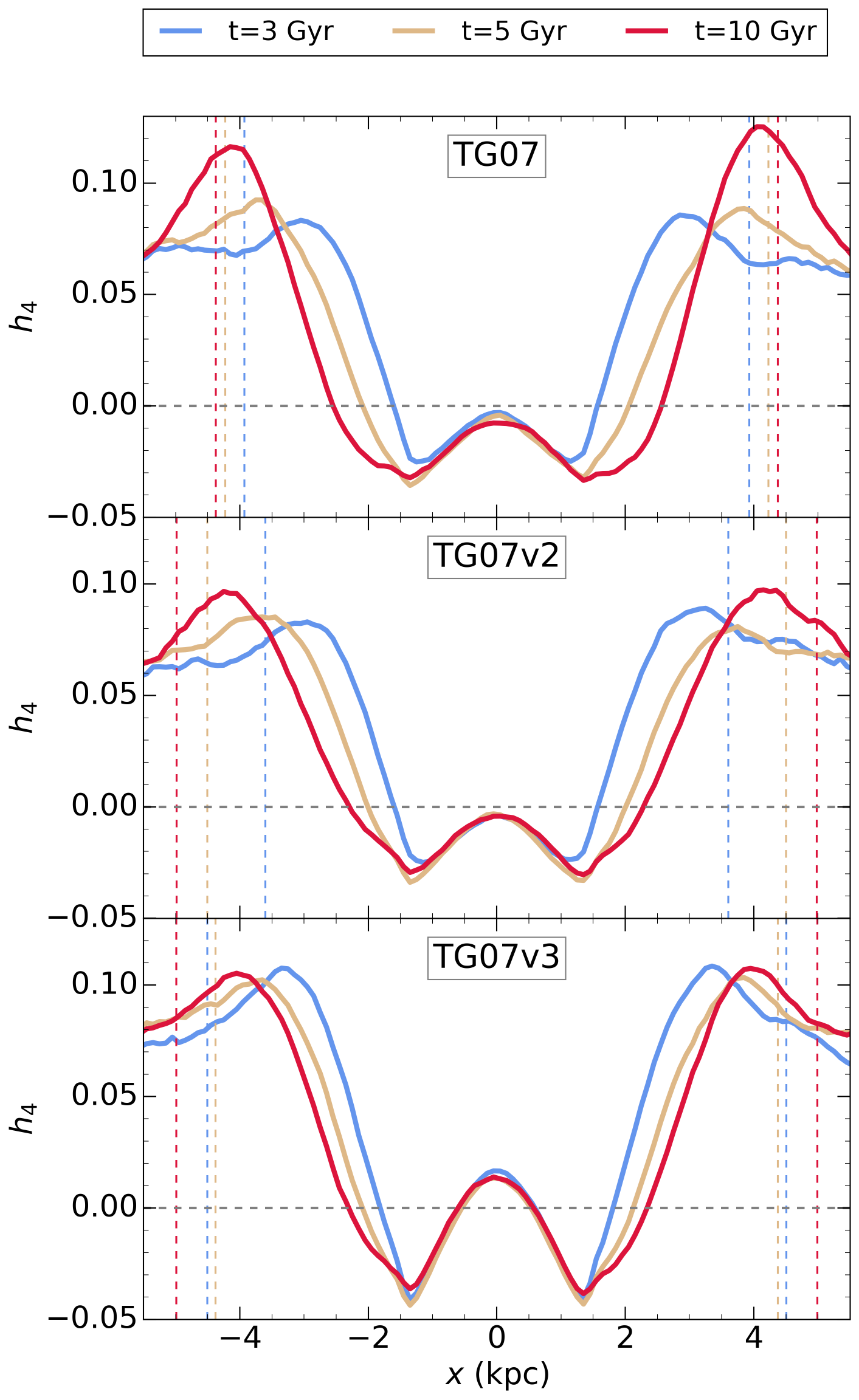}
\caption{The evolution of the $h_4$ profiles along the bar's major axis in the three TG07 models. The double minima are a signature of the BP bulge. The vertical dashed lines indicate the radius of the bar at the given time, while the horizontal dashed line indicates $h_4=0$.}
  \label{Fig:TG07h4s}
\end{figure}

One method of quantifying the box-peanut shape of a bulge is with the fourth order Gauss-Hermite moment, $h_{4}$ (see \citealt{Debattista2005}) of the vertical velocity distribution along the bar's major axis.  The presence of a BP bulge is revealed by the presence of a double minimum in the $h_4$.  Figure \ref{Fig:TG07h4s} shows the time evolution of the $h_{4}$ moment across the TG07 family of models.  The similarities of the models indicates that the BP only grows modestly between 3 Gyr and 10 Gyr, as does the bar itself. The depth of the $h_4$ minima does not change very much. We conclude that it must be the broadening of the bar that causes the change in the apparent distribution of red clump magnitudes rather than any significant weakening of the BP itself.  This result complements the work of \citet{McClure2025}.  They examined a suite of pure N-body models with differing classical bulge fractions and found that all their models formed a BP bulge.  In their work, the BP structures form via resonances with the bar. When orbits cross the bar's horizontal and vertical resonances, especially at resonance overlaps, the BP bulge grows.  While we have not performed such detailed orbit analysis here \citep[but see also][]{Beraldo2023}, it is suggestive that a similar mechanism is operating in the TG07 models. 

Based on the TG07 models' consistency with the bar length, pattern speed, as well as its similarity to the \citet{Gonzalez2015} line-of-sight distributions, we argue that it has produced a plausible MW bar and BP-bulge. In Appendix \ref{MWComp} we further compare the TG07 simulation to other MW observational constraints.  The original \citet{TepperGarcia2021} model, on which the ICs of model TG07 are based, was tuned to match the observed rotation curve and surface density profile of the MW.  As the TG07 model evolves, it ends up moving away from these conditions.  However, model TG07 remains equally consistent (or inconsistent) with MW observations as the original \citet{TepperGarcia2021} model. 

It is worth discussing briefly the dynamical evolution of tailored simulations and comparisons to the MW (or other systems).  While the TG07 model is fairly stable, some aspects of it do continue to evolve with time.  If the MW bar's pattern speed is indeed declining, as suggested by \citet{Chiba2021}, then a simulation can only match the MW's bar properties for a period of $\sim1$ Gyr. Moreover, changes in the mass distribution will be reflected in the rotation curve, surface density profile, and other observations.  Thus, a `successful' simulation may only agree with the full set of available MW measurements for a relatively short period of time.  To move forwards in this regime, it will be necessary to build a suite of simulations that are designed to evolve towards MW observations (rather than starting with ICs that match MW observations).  A successful model in such a suite will only match these observations for a brief period of time before evolving away.  This prospect of `snapshot' matching then opens up an interesting regime where it would be possible to date specific structures seen by the time it takes for the model to evolve to their current observed configuration.  Such an effort is beyond the scope of this work, where we are focused on building a plausible model of the MW rather than precisely matching it.  In that sense, the TG07 model is indeed plausible, meaning that it is possible to build a realistic MW model that forms a BP bulge that is similar to observations from an initially bulge-less disc.


\section{Conclusions}\label{Sec:Conclusions}

In this work we generated a plausible model for the MW by evolving a system that initially comprised a S\'{e}rsic disc, a gas disc, and a dark halo. Though all of the simulated systems formed bars and BP bulges, TG07 was best able to reproduce observational data for the length, strength, pattern speed, and line-of-sight density.  All simulations produce a bar and BP bulge.  However, the bars and BPs in the gasless models grow far too strong for the MW, due to the very strong secular growth of the bar.  The inclusion of even a modest gas disc, of only 7\% of the stellar mass, slows down the secular evolution of the bar, consistent with both the findings of \citet{Beane2023} and \citet{Athanassoula2013}.  In particular, \citet{Athanassoula2013} found that the gas has a dual effect of both preventing bar formation, and, when bars form, causing it to evolve more slowly.

The TG07 models evolve slowly, with a metastable bar and BP structures.  The stability of the bars is due to the presence of the gas as differences in the random seeds or feedback receipes all produce stable models.  The TG07 bars all match the observed properties of the MW bar, with the exception of the slow down rate found by \citet{Chiba2021}, though this was based on the assumption that the pattern speed was a linear function of time, which is not the case in the TG07 models. 

In addition to matching the majority of the bar properties, the BP-bulge structures of the TG07 simulations broadly and qualitatively reproduce the overall distribution of red clump and red giant branch bump stars seen along various lines-of-sight in the MW \citep{Gonzalez2015}.  Moreover, other observations, such as the rotation curve and surface density profile, are also consistent with MW observations.  Thus, it is indeed possible to build a plausible MW bar and BP~bulge using bulgeless ICs, provided that a gas disc is present.

While there is significant observational evidence that the MW has an in-situ bulge formed through the secular evolution of the disc \citep{Shen2010,Debattista2017,KormendyBender2019}, it remains unclear whether such bulges are common in $\Lambda$CDM.
\citet{Governato2010} used cosmological simulations to show that it is possible to generate nearly exponential discs in dwarf galaxies which fail to produce a bulge via supernova feedback acting on an inhomogeneous interstellar medium. In more massive galaxies, such as the MW, therefore it is conceivable that similar processes can produce similarly bulgeless galaxies with S\'ersic profiles with denser centres. Further study of modern cosmological simulations will be required to confirm that MW-mass galaxies can indeed form without a classical bulge.

Ultimately, our ability to generate a plausible MW using bulgeless ICs opens new avenues of exploration. We are now able to explore BP~bulge formation over a large parameter space and can perform a simulation campaign to find the best possible models of the Galaxy.  Similar experiments can be performed for other galaxies with observed BP~bulges. High resolution extensions of these simulations can be used to study the phase-space structure and compare it to those observed with {\it Gaia}.  There are many other experiments that can be run to explore bulgeless ICs, and our new version of \galactics\ provides the key tool needed.  


\section*{Acknowledgements}
We thank the referee, Elena D'Onghia, for her helpful comments and suggestions.
The simulations in this paper were run at the DiRAC Shared Memory Processing system at the University of Cambridge, operated by the COSMOS Project at the Department of Applied Mathematics and Theoretical Physics on behalf of the STFC DiRAC High Performance Computing (HPC) Facility: www.dirac.ac.uk. This equipment was funded by the Department for Business, Innovation and Skills (BIS) National E-infrastructure capital grant ST/J005673/1, STFC capital grant ST/H008586/1, and STFC DiRAC Operations grant ST/K00333X/1. DiRAC is part of the National E-Infrastructure. LMW was supported by a Discovery grant through the Natural Sciences and Engineering Research Council of Canada.


\section*{Data availability}

The data from the suite of simulations is available upon request.  The \galactics\ code is publicly available via GitHub (see footnote 1).




\bibliographystyle{mnras}
\bibliography{SersicDisk}

\appendix

\section{\textsc{GalactICS} with a S\'{e}rsic disc}\label{sec:Galactics}

The \galactics\ code is a robust method of generating ICs for tailored simulations.  While initially designed to generate collisionless systems \citep{Kuijken1995,Widrow2005,Widrow2008}, \citet{Deg2019} modified the code to include an exponential gas disc.  Modifying this version of the code to incorporate S\'{e}rsic stellar discs is relatively straightforward due to the nature of the disc density-potential pair used in the code.  The disc density and potential are written as 
\begin{equation}
    \rho_{d}(R,z)=\rho_{hh}(R,z)+\rho_{r}(R,z),
\end{equation}
\begin{equation}
     \Phi_{d}(R,z)=\Phi_{hh}(R,z)+\Phi_{r}(R,z). 
\end{equation}
where $\rho_{hh}-\Phi_{hh}$ form an analytic density-potential pair that captures the high harmonics of the disc while $\rho_{r}$ and $\Phi_{r}$ are the residual density and potential, which are approximated by a Legendre polynomial series. The disc density is taken to be
\begin{equation}\label{Eq:Gendisc}
    \rho_{d}(R,z)=\Sigma(R,R_{d})f(z,z_{d})
    C\left(R,R_{d},R_{t,d},\delta R_{t} \right),
\end{equation}
where $\Sigma(R,R_{d})$ is the disc surface density, $f(z,z_{d})$ is the vertical profile, and $C\left(R,R_{d},R_{t,d},\delta R_{t} \right)$ is a truncation function that smoothly sets the disc density to zero at the disc truncation radius, $R_{t,d}$ over a width of $\delta R_{t}$.
In \textsc{GalactICS}, the vertical profile of a stellar disc is given by
\begin{equation}
f(z,z_{d})=\textrm{sech}^{2}\left(\frac{z}{z_{d}} \right),
\end{equation}
where $z_{d}$ is the disc vertical scale height.  The $\textrm{sech}^{2}$ function is often used to describe stellar discs \citep{Kuijken1995} as it integrates easily.  For \galactics\ a $\textrm{sech}^{2}$ vertical profile is particularly helpful for the analytic density-potential pair.  For a S\'{e}rsic disc, the disc surface density is 
\begin{equation}
\Sigma(R,R_{d})=\Sigma_{0}\textrm{exp}\left(
    -\left(\frac{R}{R_{d}}\right)^{1/n}\right)~,
\end{equation}
where $\Sigma_{0}$ is the scale density, $R_{d}$ is the disc scale length, and $n$ is the S\'ersic index.

The high harmonic disc density needed to modify \galactics\ to generate a S\'ersic disc can be constructed by generalizing the high harmonic pairs used for the exponential disc in previous versions of \galactics\  \citep{Deg2019,Widrow2008,Widrow2005,Kuijken1995}.  The generalized potential is set to
\begin{equation}
    \Phi_{hh}(R,z)=-2 \pi \Sigma(r)f(z)z_{d}~,
\end{equation}
where $r$ is the spherical radius.  Then using Poisson's equation, the corresponding high harmonic density is given by:
\begin{equation}
    \begin{split}
    \frac{\rho_{hh}(R,z)}{2}=\frac{d^{2}\Sigma(r)}{dr^{2}}f(z)z_{d}
    +2 \frac{d\Sigma(r)}{dr}f(z)z_{d} \\
    +2\frac{d\Sigma(r)}{dr}\frac{df(z)}{dz} z
    +\Sigma(r) \frac{d^{2}f(z)}{dz}\frac{1}{z_{d}}
    \end{split}
\end{equation}
where $\Sigma(r)$ is the surface density calculated using the spherical radius and $f(z)$ is the vertical profile.  All that is then required is calculating the various derivatives for the S\'{e}rsic surface density and $\textrm{sech}^{2}$ vertical profile.  With the high harmonic terms in hand, the full density and potential pair can be calculated by adding these terms to the residual terms calculated from the Legendre polynomials. 

In order for the model to be in equilibrium, it is necessary to set the velocities of each particle.  As in the \citet{Deg2019} version of \galactics\ the vertical velocity dispersion profile is set by the disc thickness.  However, the S\'{e}rsic disc requires an expansion of the exponential profile previously used for the radial and tangential velocity dispersions.  At $n>1$, S\'{e}rsic discs are more centrally concentrated than exponential discs, which can cause the underlying assumptions of the \galactics\ DF to no longer hold.  To address this issue, we replace the single exponential dispersion profile with a double exponential:
\begin{equation}
    \sigma_{R}^{2}(R)=\sigma_{1}^{2}e^{-R/R_1}+\sigma_{2}^{2}e^{-R/R_2}~.
\end{equation}

This approach decouples the radial velocity dispersion profile from the underlying density, but it allows for more realistic galaxy ICs.


\section{Further comparisons with the MW}\label{MWComp}

\begin{figure}
\centering
    \includegraphics[width=0.4\textwidth]{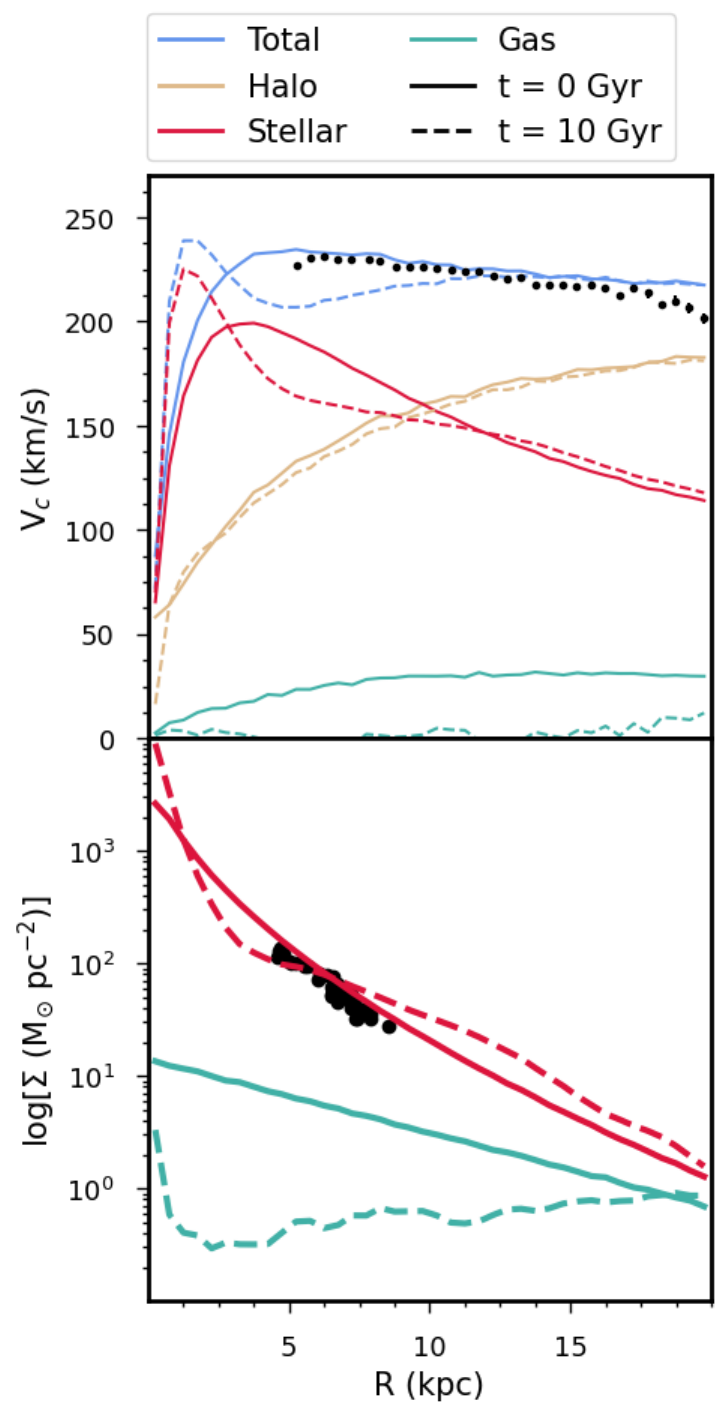} 
    \caption{The initial (solid lines) and final (dashed lines) rotation curve and surface density profiles for the TG07 model (both stellar and gaseous).  The rotation curve data is from \citet{Eilers2019} and the surface density data are from \citet{Bovy2013}. }
  \label{Fig:TepperGarcia_RC_SD}
\end{figure}

The TG07 simulation presented in Sec. \ref{SSec:TG07Sims} generates a bar and BP-bulge comparable to the MW's.  We therefore compare other properties of the system to the MW.  The upper panel of Figure \ref{Fig:TepperGarcia_RC_SD} compares the rotation curve  of the $t=0$ and $t=10~\rm{Gyr}$ snapshots to the measured RC of \citet{Eilers2019}.  Given that the TG07 model has ICs that are based on \citet{TepperGarcia2021}, who designed their simulation to match the RC of \citet{Eilers2019}, it is unsurprising that model TG07 initially matches the \citet{Eilers2019} observational data.  As the system evolves and the bar develops, the TG07 simulation moves away from the \citet{Eilers2019} RC.  Between $R=5-10~\rm{kpc}$, the $t=10$~Gyr snapshot of model TG07 has a RC that is $\sim15-20~\kms$ lower than in the MW.  More importantly, the model RC is rising in this regime while the \citet{Eilers2019} RC is decreasing. It is worth noting that the original \citet{TepperGarcia2021} simulation shows a similar change to the RC with an equal discrepancy to the \citet{Eilers2019} data between $R=5-10~\rm{kpc}$.

\citet{TepperGarcia2021} also designed their initial model to reproduce the MW's total surface density profile (which includes both the stellar and dark matter densities) measured by \citet{Bovy2013}.  \citet{TepperGarcia2021} assumed equal contributions from stars and dark matter in the \citet{Bovy2013} surface density (which covers a radial range of $\sim5-9~\rm{kpc}$).  The bottom panel of Figure \ref{Fig:TepperGarcia_RC_SD} shows the comparison of the TG07 simulation to the \citet{Bovy2013} SD profile (with that same factor of two).  As with the RC, the TG07 ICs match observations of the MW, but by $t=10~\rm{Gyr}$ the model has moved away from the starting SD.  However, it is worth noting that both the \citet{Bovy2013} and  TG07 model at $t=10~\rm{Gyr}$ show an inflection in the SD profile, but in the case of model TG07, there are two distinct inflections; one from a steep inner slope due to the BP-shaped bulge to a flatter slope around $R\sim3~\rm{kpc}$ and a second, less extreme, one to a steeper slope around $R\sim 7~\rm{kpc}$.

\begin{figure}
\centering
    \includegraphics[width=0.5\textwidth]{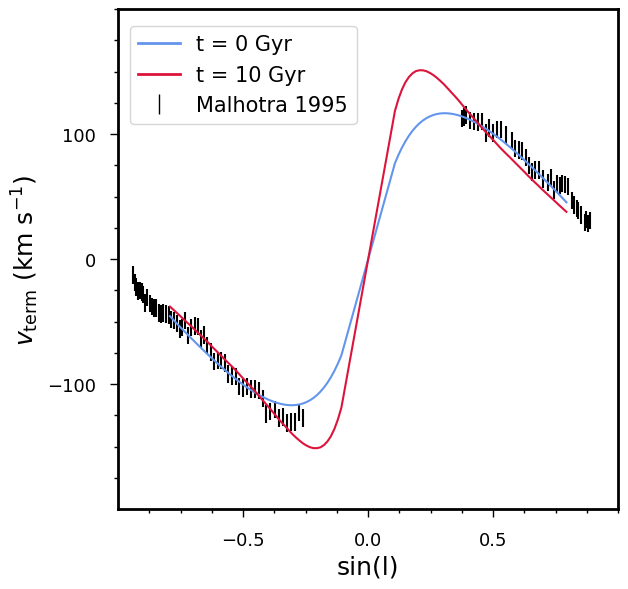} 
    \caption{The initial and final terminal velocity curves of the TG07 model.  The data points are drawn from the terminal velocities of \citet{Malhotra1995}.}
  \label{Fig:TepperGarcia_TV}
\end{figure}

The \citet{Eilers2019} RC covers the outermost regions of the MW but the inner RC is often probed by the terminal velocity curve. Figure \ref{Fig:TepperGarcia_TV} explores this inner RC by comparing the TG07 simulation to the terminal velocity observations of \citet{Malhotra1995}.  At $t=0$, the TG07 model is, as expected from the RC shown in \ref{Fig:TepperGarcia_RC_SD}, consistent with observations.  As the bar and BP~bulge form there is a significant rearrangement of material in the inner region, which is reflected in the larger amplitude of the innermost terminal velocities at $t=10~\rm{Gyr}$ seen in Figure \ref{Fig:TepperGarcia_TV}.  Nonetheless, the evolved TG07 model remains consistent with the \citet{Malhotra1995} data.

Beyond these, it is possible to compare the TG07 model to local MW observations, including the Oort constants, local circular speed, and local surface density.  These comparisons are listed in Table \ref{tab:MWMeas} for the $t=0$, 5, and 10 Gyr TG07 snapshots.  There is clearly more work to be done to precisely match these local constraints, particularly the Oort constants and local circular speed.  However, given that the model is not tuned to these constraints, the evolved TG07 model has local measurements that are remarkably similar to MW observations.

\begin{table*}
    \centering
    \begin{tabular}{|c|c|c|c|c|}
        \hline
         Parameter & Measured & TG07 - 0 Gyr & TG07 - 5 Gy & TG07 - 10 Gyr \\
         \hline 
         $v_{lsr}$ (km/s) & 218 $\pm$ 6  \citep{Bovy2012} & 224 & 211 & 205\\
         A (km/s/kpc) & 14.8 $\pm$ 0.8  \citep{Feast1997}& 13.6 & 11.8 & 10.2\\
         B (km/s/kpc) & -12.4 $\pm$ 0.6 \citep{Feast1997} & -13.7 &-14.1& -14.9 \\
         $\Sigma$ (\msol \textrm{pc}$^{-2}$) & 49 $\pm$ 9 \citep{Flynn1994} & 38.7 & 51.8 & 46.2 \\
         \hline 
    \end{tabular}
    \caption{Local measurements of the MW compared to the model TG07 at 3 epochs. The rows, from top to bottom, are the local standard of rest, $v_{lsr}$, the Oort A and B constants, and the local surface density.}
    \label{tab:MWMeas}
\end{table*}

\begin{figure*}
\centering
    \includegraphics[width=0.8\textwidth]{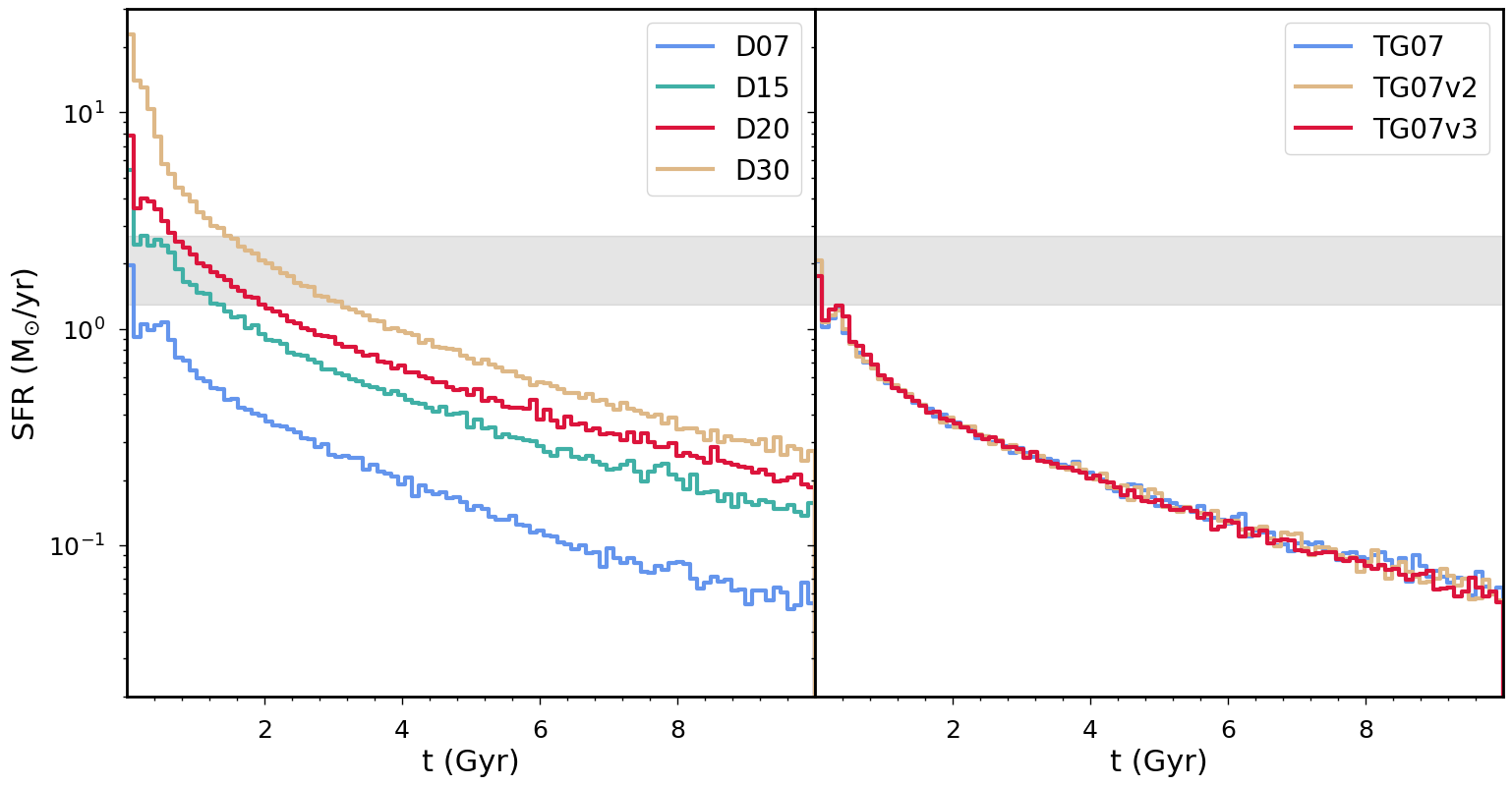} 
    \caption{The SFR for the D-family (left) and TG-family (right) of simulations that include gas discs. The grey shaded region shows the SFR measured in the MW \citep{Elia2022}.
    }
  \label{Fig:SFR}
\end{figure*}

In addition, we have also calculated the star formation rate (SFR) of the simulation.  Figure \ref{Fig:SFR} shows the SFR of all simulations that include gas discs as well as the measured value of $2.0\pm0.7~\rm{M}_{\odot}~\rm{yr}^{-1}$ for the MW \citep{Elia2022}.  In all simulations, the SFR is substantially below the \citet{Elia2022} value, due to the fact that there is no replenishment of the gas reservoir in any of the runs.  This issue will arise in any tailored simulation that does not include some form of gas accretion.  In all simulations, there is an initial spike in the SFR at $t=0~\rm{Gyr}$ due to the \galactics\ gas disc initialization, after which it declines with time.   

\section{Metastability}
\label{app:metastability}

\begin{figure*}
\centering
    \includegraphics[width=0.8\textwidth]{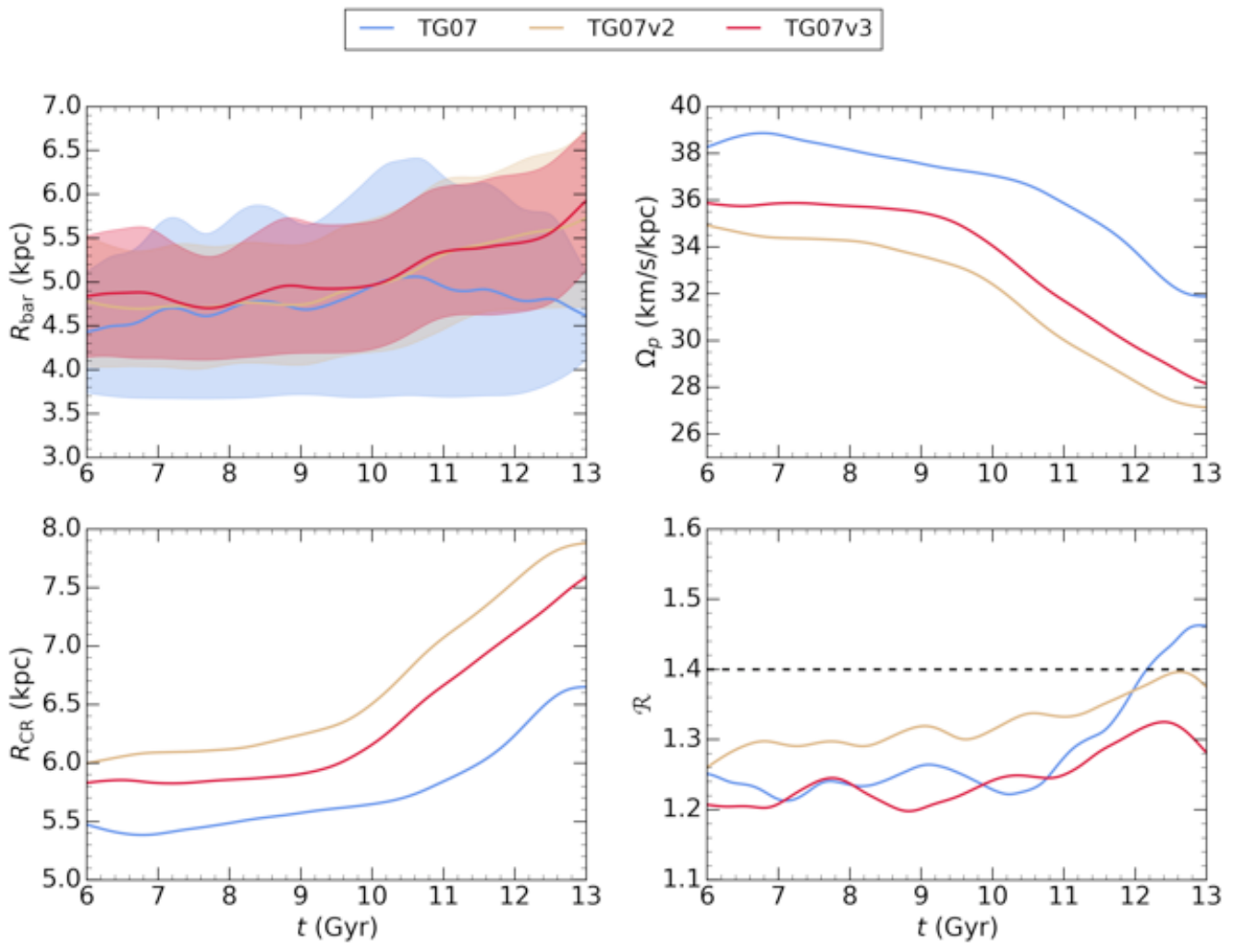} 
    \caption{Corotation analysis for the gas models TG07, TG07v2 and TG07v3. Each plot shows the evolution of the respective variable with time, and has been smoothed for clarity. From upper left to bottom right: the bar radius $R_\mathrm{bar}$, the bar pattern speed as computed at each time step using the algorithm of \citet{Dehnen+2023}; the corotation radius $R_\mathrm{CR}$, and $\mathcal{R} = R_\mathrm{CR}/R_\mathrm{bar}$. In the bottom right panel, the canonical division between fast and slow bars ($\mathcal{R}=1.4$) is indicated by the horizontal black dashed line. TG07v2 and TG07v3 host fast bars for almost all their evolution.}
  \label{Fig:TG07_metastable}
\end{figure*}

We have interpreted the steady evolution of the bar in the TG07 models as being due to the metastability identified by \citet{sellwood_debattista06}. This metastable state arises because the pattern speed is briefly forced to rise, which traps the bar into facing resonances with {\it increasing} phase space density. The rising pattern speed can easily be induced in simulations such as those presented here by bars funneling gas inwards, which accounts for why a small fraction of gas is able to have such a strong effect. \citet{sellwood_debattista06} argued that small perturbations, such as those from halo substructure, are able to return the bar to a steady evolution. In the absence of such perturbations, \citet{sellwood_debattista06} suggested that the bar eventually leaves the metastable state due to secular evolution at higher order resonances. As we have seen, during the metastable state when the bar pattern speed is more or less steady, the bar is still evolving (growing wider) suggesting that such secular evolution is still active in the background. In Figure~\ref{Fig:TG07_metastable} we show that, when evolved further, the bars in the TG07 models do indeed leave the metastable state. In all three models we find that by 13 Gyr the bar is once again slowing (upper right panel). In models TG07v2 and TG07v3 the bar size is rising rapidly, whereas in model TG07 the bar size remains roughly constant (upper left panel). Moreover, the corotation radius is increasing for all models (lower left panel), with TG07 and TG07v2 both crossing the limit of $\mathcal{R}=1.4$ to become slow rotators (lower right panel).  This variation in bar evolution is reflective of the stochasticity expected in such cases \citep{sellwood_debattista09}. We conclude that gas gives rise to a prolonged case of metastability in which the bar fails to grow and remains fast. 

\citet{sellwood_debattista06} also found that the metastable states in their models were quite sensitive to minor perturbations, whether from a massive orbiting particle meant to mimic a satellite or small disc perturbations. In light of this, it seems not unlikely that the metastable state is quite fragile, and liable to be broken by external perturbations, such as would arise in a fully cosmological setting. Confirmation needs further simulations with substructure taken into account.

\section{Supplemental Figures}
\label{app:supplemental}

\begin{figure*}
\centering
    \includegraphics[width=\textwidth]{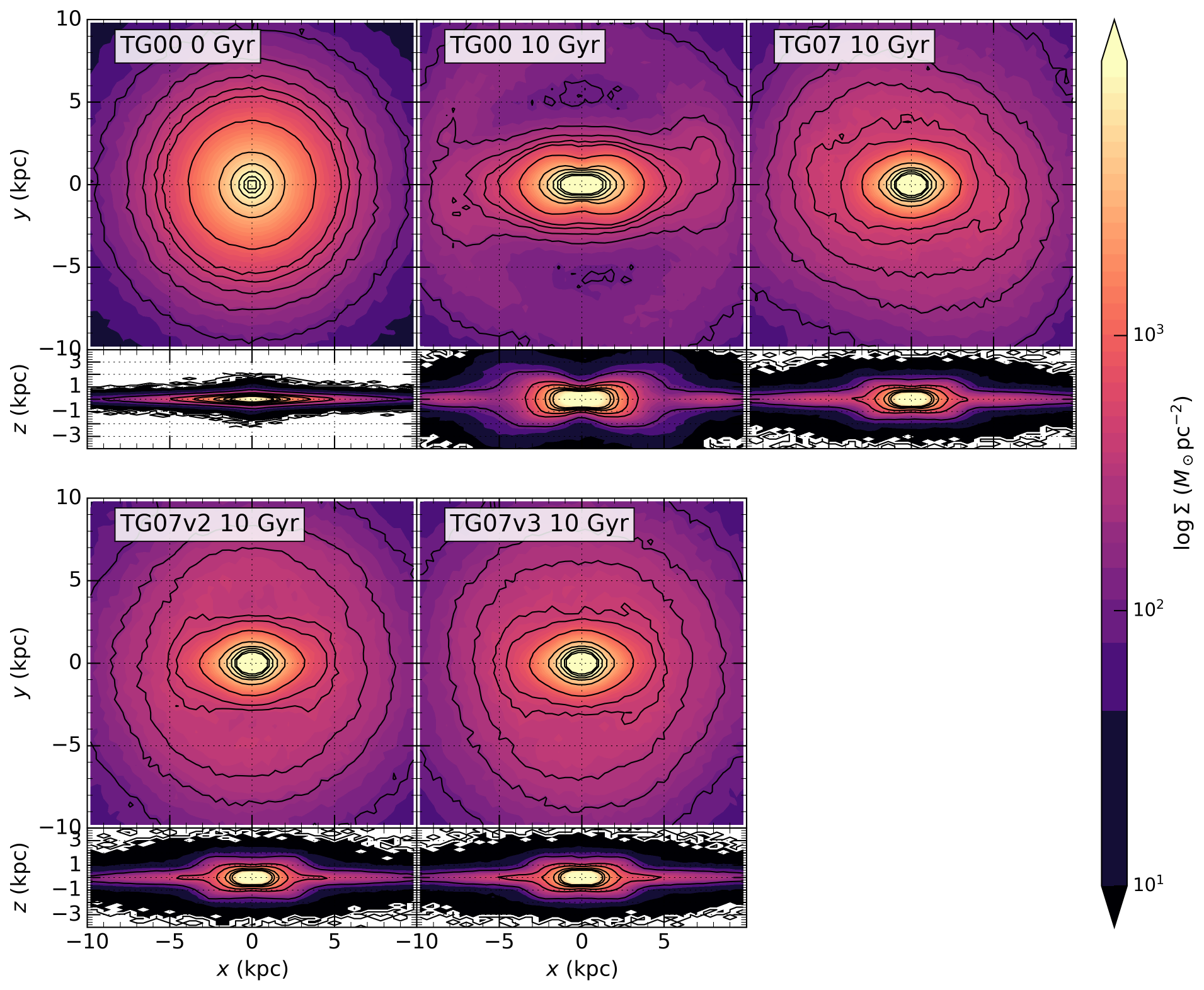} 
    \caption{A comparison of the final outcome of the TG-sequence of models. For the $(x,z)$-plane views, we have imposed a cut $|y|<1$ kpc to emphasize the BP nature of the bulge.}
  \label{Fig:TG_SD_Fin}
\end{figure*}

\begin{figure*}
\centering
    \includegraphics[width=0.8\textwidth]{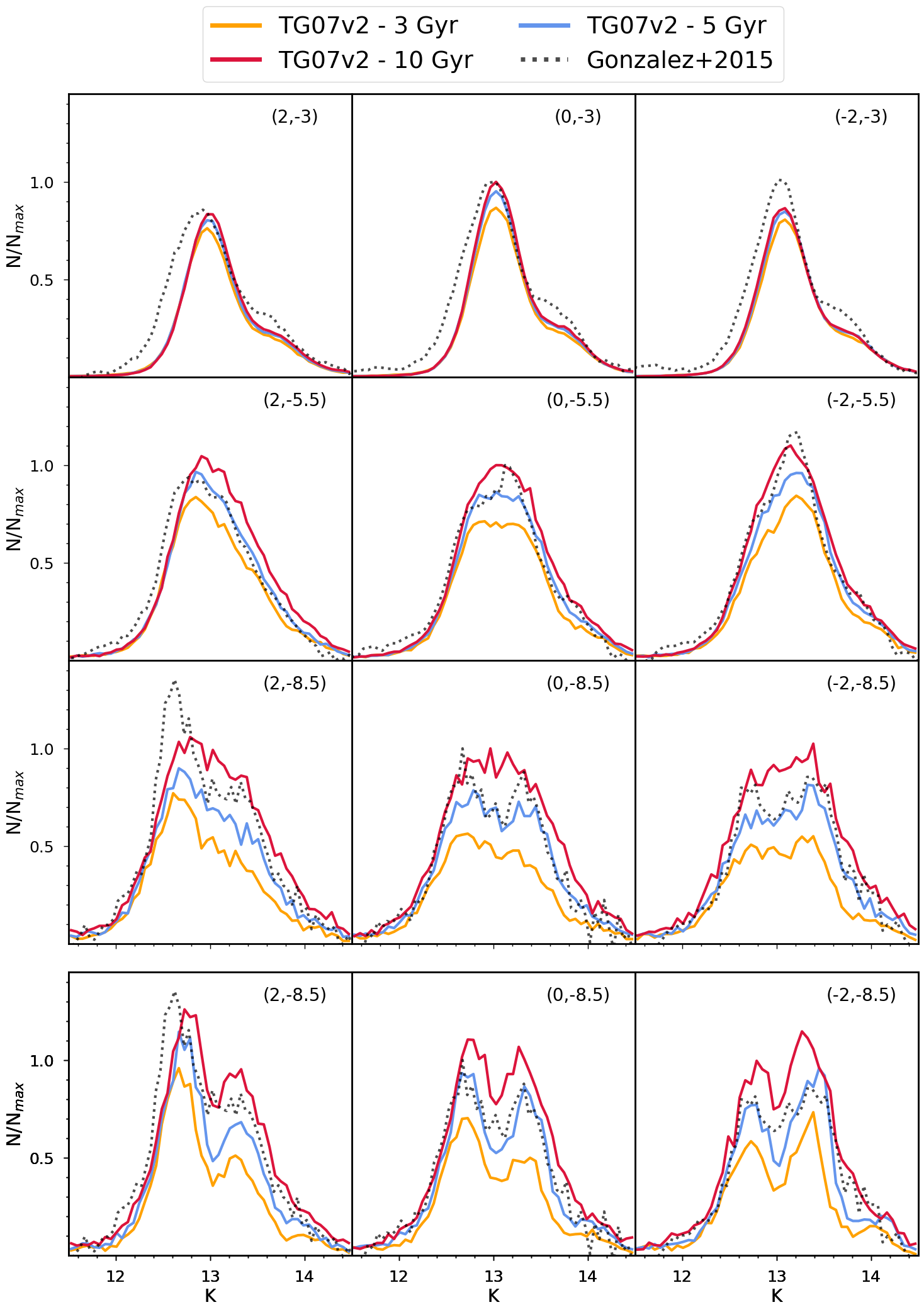} 
    \caption{
    Similar to Figure~\ref{Fig:DOnghiaBulgeLOS} for model TG07v2.}
  \label{Fig:TG07v2_BPshape}
\end{figure*}

\begin{figure*}
\centering
    \includegraphics[width=0.8\textwidth]{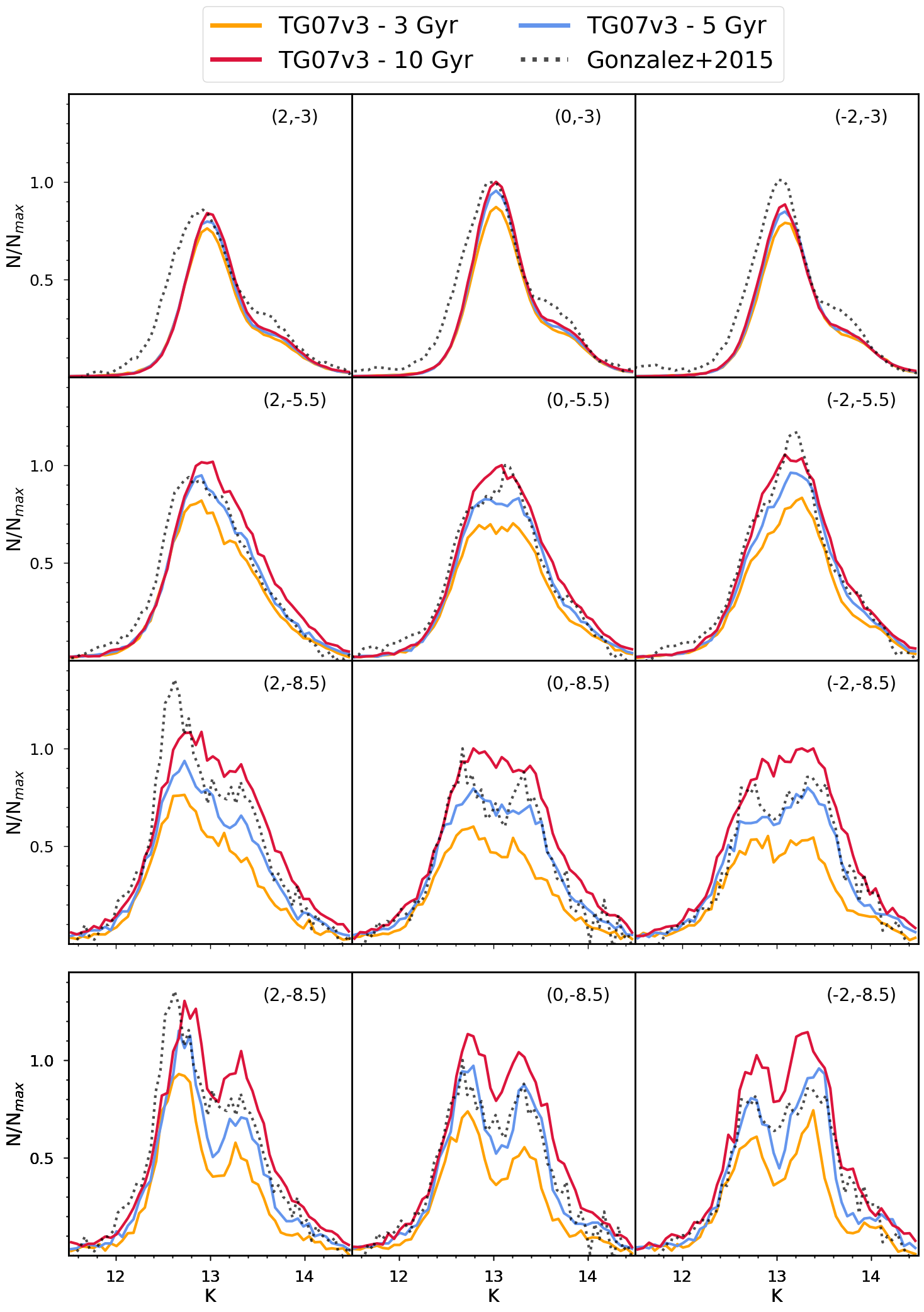} 
    \caption{Similar to Figure~\ref{Fig:DOnghiaBulgeLOS} for model TG07v3.
    }
  \label{Fig:TG07v3_BPshape}
\end{figure*}

For completeness, we include a set of supplemental figures showing other aspects of the evolution of the models.  Figure \ref{Fig:TG_SD_Fin} shows the TG family of models at $t=10$ Gyr similar to Figure \ref{Fig:DOnghiaStellarSDMaps}.  Figures \ref{Fig:TG07v2_BPshape} and \ref{Fig:TG07v3_BPshape} shows the time evolution of the bulge line-of-sight observations compared to \citet{Gonzalez2015} similar to Figure \ref{Fig:TepperGarcia_BulgeLOS}.

\begin{figure*}
\centering
    \includegraphics[width=0.8\textwidth]{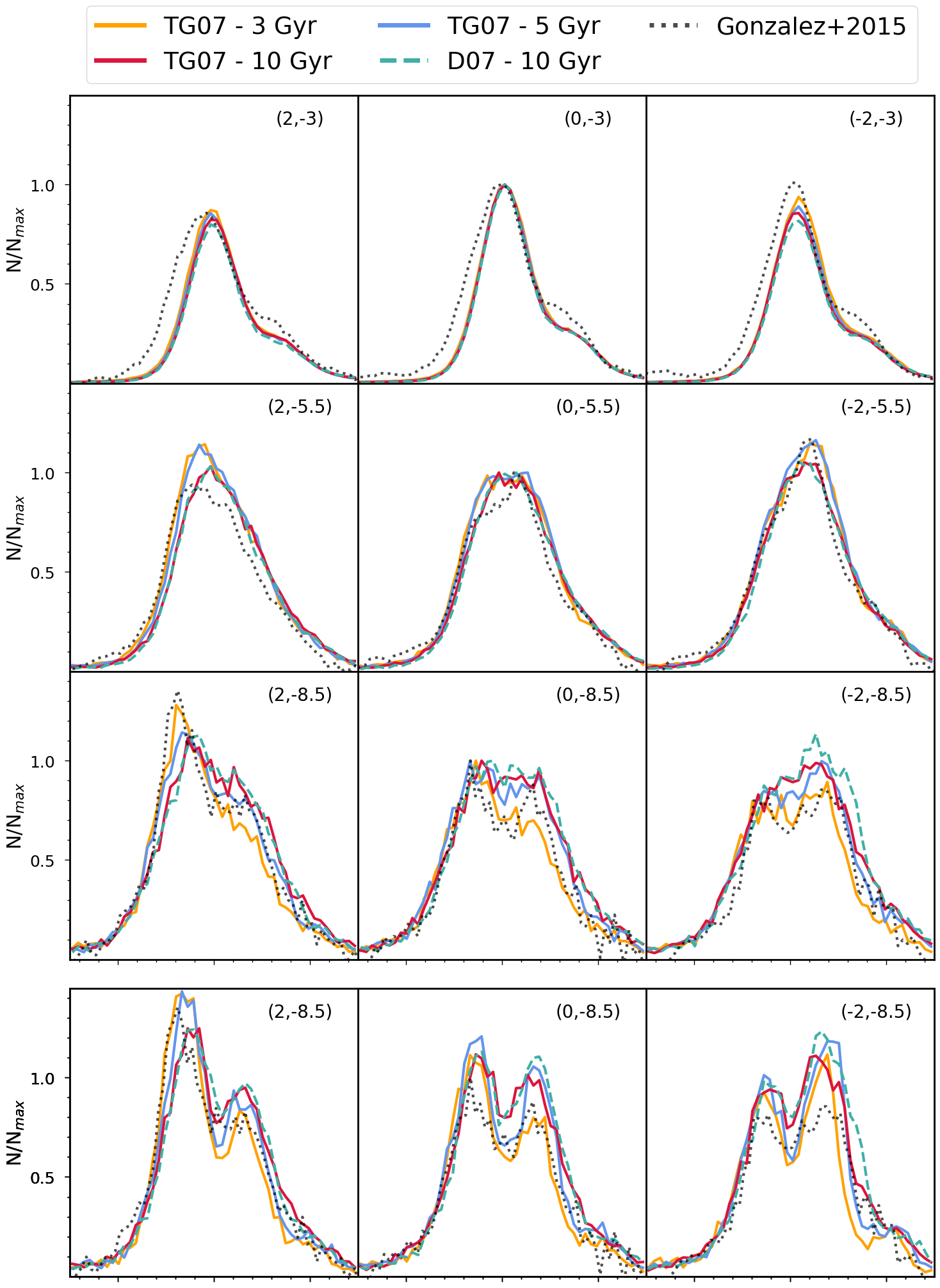} 
    \caption{Similar to Figure~\ref{Fig:TepperGarcia_BulgeLOS} except each curve is normalized individually to their own peak in the $l=0^{\circ}$ panels (rather than normalizing to the singular peak).  Additionally, this plot includes the D07 model at $t=10$ Gyr for comparison.
    }
  \label{Fig:TG07_D07_AltNorm}
\end{figure*}

Figure \ref{Fig:TG07_D07_AltNorm} requires additional commentary as it uses an alternate normalization than Figures \ref{Fig:DOnghiaBulgeLOS} and \ref{Fig:TepperGarcia_BulgeLOS}.  In this Figure, the bulge line-of-sight curves are individually normalized to their peaks in the $l=0^{\circ}$ panels.  This normalization is more suitable for direct comparisons to the \citet{Gonzalez2015} data.  With this normalization Figure \ref{Fig:TG07_D07_AltNorm} shows that the D07 and TG07 models at $t=10$ Gyr have similar shapes.  It is also clear that there are features in the \citet{Gonzalez2015} data that are missed by the various snapshots.  For example, the $(0^{\circ},-5.5^{\circ})$ panel has a single peak at $K>13$, while all the snapshots have a double peak.  This result highlights a few key points.  While the TG07 model results in a BP bulge with similar features to those observed in the MW, it is not a perfect match, which is consistent with the RC and SD results shown in Figure \ref{Fig:TepperGarcia_RC_SD}.  Detailed matching will require a larger simulation campaign as well as a quantification of the uncertainties in these bulge line-of-sight measurements.

\bsp	
\label{lastpage}
\end{document}